\documentclass{aa}
\usepackage{epsfig}
\usepackage{graphicx}
\begin{document}

\title{Old stellar Galactic disc in near-plane regions according to 2MASS:
scales, cut-off, flare and warp}

\author{
M. L\'opez-Corredoira\inst{1,2}, A. Cabrera-Lavers\inst{2}, 
F. Garz\'on\inst{2,3}, P. L. Hammersley\inst{2}}
\institute{$^1$ Astronomisches Institut der Universit\"at Basel,
Venusstrasse 7, CH-4102 Binningen, Switzerland\\
$^2$ Instituto de Astrof\'\i sica de Canarias, E-38200 La Laguna, 
Tenerife, Spain \\
$^3$ Departamento de Astrof\'\i sica, Universidad de La Laguna, Tenerife, Spain
}

\offprints{martinlc@astro.unibas.ch}

\date{Received xxxx / Accepted xxxx}

\abstract{
We have pursued two different methods to analyze the old stellar
population near the Galactic plane, using data from  the 2MASS survey.
The first method is based on the isolation of the red clump giant
population in the color-magnitude diagrams and 
the inversion of its star counts to obtain directly
the density distribution along the line of sight. The second method
fits the parameters of a disc model to the star counts
in 820 regions.
Results from both independent methods are consistent with each other.
The qualitative conclusions are that the disc is well fitted by
an exponential distribution in both the galactocentric distance and height.
There is not an abrupt cut-off in the stellar disc (at least within 
$R<15$ kpc). There is a strong flare (i.e.
an increase of scale-height towards the outer Galaxy) which 
begins well inside the solar circle, and hence there 
is a decrease of the scale-height towards the inner Galaxy.
Another notable feature is the existence of a warp in the old stellar
population whose amplitude is coincident with the amplitude of the gas warp.\\
It is shown for low latitude stars
(mean height: $|z|\sim 300$ pc) in the outer disc (galactocentric radius 
$R> 6$ kpc) that: the scale-height in the solar
circle is $h_z(R_\odot)=3.6\times 10^{-2}R_\odot$, the scale-length of the
surface density is $h_R=0.42R_{\odot}$ and  the scale-length
of the space density in the plane (i.e. including the effect of the
flare) is $H=0.25R_{\odot}$. The variation of the scale-height
due to the flare follows roughly a law $h_z(R)\approx  h_z(R_\odot)
\exp \left(\frac{R-R_\odot}{[12-0.6R({\rm kpc})]\ {\rm kpc}}\right)$ 
(for $R<\sim 15$ kpc; $R_\odot=7.9$ kpc). 
The warp moves the mean position of the disc to a height 
$z_w=1.2\times 10^{-3} R({\rm kpc})^{5.25}\sin (\phi +5^\circ )$ pc 
(for $R<\sim 13$ kpc; $R_\odot=7.9$ kpc).
}
\titlerunning{Disc/2MASS}

\maketitle

\section{Introduction}

Star counts have been widely used for the study of Galactic structure (see Paul
1993), and are becoming increasingly important with the appearance of 
wide area surveys in the
last decades (see for example Bahcall 1986, Majewski 1993, Reid 1993, Garz\'on et al. 1993,
Price 1988). The use of progressively larger and more sensitive
samples  combined with detailed models of stellar galactic distribution (for example Bahcall \&
Soneira, 1980, Wainscoat et al. 1992) has overcome most of the original
uncertainties and difficulties in the analysis and interpretation of star count
data. The  advent of NIR detector arrays has permitted
the detailed exploration of the stellar structure in hitherto hidden areas of the Milky
Way, such as the Galactic Plane and Bulge (Eaton et al. 1984, Habing 1988, 
Garz\'on et al. 1993, Hammersley et al. 1994, Ruphy et al. 1996,
L\'opez--Corredoira et al. 2000, Epchtein 1997,
Skrutskie et al. 1997).  This is because at these wavelengths 
interstellar extinction is significantly less when compared
to the visible, while keeping the individual stellar contribution to the
observed flux.

There are, however, still controversial or totally unknown parameters
in the description of the detailed stellar structure. Some of these
are concerned with the radial and vertical distribution of the Galactic
disc, and its specific morphology. Several radial scale-lengths have
been determined by different groups, even when  using the same data
sets. As a general trend, the shorter scale-lengths are found for the
surveys using longer wavelengths (Kent et al. 1991), although it
could equally be that different source type or regions of the Galaxy
are being examined.  There is far less debate on the vertical
scale-height, with the canonical value for the old population in the
Solar vicinity being about 300 pc. Still more work is needed, however,
to differentiate between the distribution being an exponential  or the
$sech^2$ (van der Kruit 1988). Finally, features in the stellar disc
like internal (Freudenreich 1998, L\'opez-Corredoira et al. 2001) or 
external (Habing 1988, Ruphy et al. 1996) cut--offs radii, 
flares (Kent et al. 1991, Alard 2000), warps 
(Djorgovski \& Sosin 1989, Freudenreich 1998,
Porcel \& Battaner 1995, Alard 2000, Drimmel \& Spergel 2001)
and/or local corrugations of the
galactic plane (Hammersley et al. 1995) are less well studied, 
and are more controversial.

In this paper we will address some of the above topics by analyzing the
shape of the stellar Galactic disc in two ways. We will first extract a
single stellar type by using the deep NIR color--magnitudes diagrams
(CMD) from the 2MASS survey (Skrutskie et al. 1997). The analysis will
be based on the red clump giants, whose relatively high density and  bright
absolute magnitude  give rise to an identifiable feature in the CMDs.
Hence,  the distance and extinction to each star can be directly
determined only assuming  the absolute magnitude and color of the
sources.  As a second approach, we will use the star counts, also taken
from 2MASS, and the results will then be  compared with those
obtained from the red clump method. Both methods will be used to examine the
scale-length, scale-height, flare, outer disc cut-off and warp. 

The data for this work has been taken from the 2nd release of the 2MASS project
(Skrutskie et al. 1997, http://www.ipac.caltech.edu/2mass/releases/docs.html).
We have made use of the available data in that
second release for near plane regions in
the outer Galaxy ($45^\circ <l<315^\circ$).

\section{Model of a non-warped Galactic disc}

In order to help to 
interpret the 2MASS star counts a simple model of the density distribution in the disc was built.

The Galactic disc is assumed to be an exponential distribution whose
star density is:

\begin{equation}
\rho (R,z)=\rho_\odot \exp \left(-\frac{R-R_\odot}{h_R}\right)
\exp\left(-\frac{|z|}{h_z(R)}\right)
\frac{h_z(R_\odot)}{h_z(R)}
\label{ro1}
,\end{equation}
\begin{equation}
h_z(R)=h_z(R_\odot)\exp \left(\frac{R-R_\odot}{h_{R,flare}}\right)
\label{flare} 
,\end{equation}
where  $\rho_\odot$ is the space density on the plane in the 
neighborhood of the Sun. 

Hence the density  falls exponentially with the galactocentric
radius $R$ (scale-length $h_R$), and with the height $z$. We have 
preferred an exponential scale-height rather than  a $sech^2$, as Hammersley et al. (1999) 
shows that a  $sech^2$ law did not give a good fit to the TMGS star 
counts near the plane.  The effect of the flare is to increase the 
scale-height of the sources in the outer Galaxy, 
hence the scale-height ($h_z(R)$) is dependent on $R$.
We have assumed that the total number of stars at a specific 
galactocentric radius remains the same as it would have been without the flare  
and  so all the flare is doing is to 
distribute the sources further from the plane. In the model
this is included by having the scale-height increase exponentially
with  galactocentric radius and then have a normalizing factor  
$\left(\frac{h_z(R_\odot)}{h_z(R)}\right)$ for the density.
Whilst this is not the first model to include a flare, it
is the first to use an exponential rather than a linear flare.

Another way of expressing the density is:

\begin{equation}
\rho (R,z)=\rho_\odot \exp \left(-\frac{R-R_\odot}{H}\right)
\exp \left(-\frac{|z|}{h_z(R)}\right)
\label{ro2}
,\end{equation}
where
\begin{equation}
H=\left(\frac{1}{h_R}+\frac{1}{h_{R,flare}}\right)^{-1}
\label{H}
,\end{equation}
i.e. $H$ is the equivalent scale-length which is 
the combination of the intrinsic scale-length
of the disc  with that of the flare. Written  in this form the equivalent
scale-length 
takes the place of the normalizing factor.   
The flare increases with $R$ and so 
removes proportionally more stars from the plane with increasing distance.
The net effect is that on the plane the star density falls far more rapidly 
with increasing galactocentric distance than the intrinsic scale-length 
predicts. Hence, $h_R$ is the scale-length of
surface density, and $H$ is the scale-length of the 
space density in the plane.

We have assumed a distance from the Sun to the Galactic center to be $R_\odot=7.9$ kpc
(L\'opez-Corredoira et al. 2000).

Therefore, this model contains only
three free parameters ($h_R$, $h_z(R_\odot)$, $h_{R,flare}$)
whose determination is carried out by minimizing the $\chi^2$ of the
fit in the following way. 

For a given set of 
parameters, the space density on the plane in the neighborhood 
of the Sun can be determined using the formula

\begin{equation}
\rho_\odot=\frac{\sum _i\frac{d_it_i}{\sigma _i^2}}
{\sum _i \frac{t_i^2}{\sigma _i^2}}
\label{rosol}
,\end{equation}
where $d_i$ is the measured density in each region, $\sigma _i$ the
poissonian errors in the counts and 
$t_i$ is the model prediction with $\rho_\odot=1$.
Hence, for each set of model parameters we calculate $\rho _\odot$
through eqn. (\ref{rosol}), and after that we calculate $\chi^2$
for that particular fit. 
Once we have calculated all the $\chi^2$ for all sets in the parameter
space, we select the one which gives a minimum $\chi^2$.
The errors are due to the errors in the counts (assumed to be
poissonian) and errors coming from the calculation.

The density equation can be further simplified on plane ($b\approx 0$), as long as  $|z|/h_z(R)<<1$ along the whole line of sight:

\begin{equation}
\rho (R,z=0)=\rho_\odot \exp \left(-\frac{R-R_\odot}{H}\right)
\label{roplano}
,\end{equation}

If the external disc were truncated, then another parameter that
could be added is the radius of the cut-off $R_{\rm cut-off}$. It will
be shown in the analysis presented here (\S \ref{.res1}), however, that
there is no evidence for a cut-off to at least $R=15$ kpc,
and so this parameter will be unnecessary in the fits. Beyond $R\approx
15$ kpc, there are very few stars detected and so very little can be said.

We have ignored the contribution of the spiral arms
since in the outer Galaxy their contribution to the star count is small 
and even then it will be at the brightest magnitudes being considered 
here, giving negligible contribution for $m_K<14.0$.

Other possible additions to the standard disc model are large scale warps and the 
presence of the thick disc. 
The warp will be considered separately in \S \ref{.warp}.

The thick disc (Gilmore \& Reid 1983; Buser et al. 1998, 1999; Ojha 2001)
will not be studied here as a separate population
since the analysis presented here is not sensitive to thick disc, as
regions in general near the plane are used where its relative contribution  
is small. It will be shown in sections
\ref{.res1}, \ref{.scounts} that the thick disc 
could constitute as much as a 10\% of the total low latitude sources
selected in this paper, so the derived 
scale-height  will be somewhat larger than would be the case for the thin
disc alone. Here we will use the term  ``old disc'' as the average structure
traced by the old disc stars, including any  thick disc component (mean height 
around 300 pc). The determined mean scale-height might be larger at higher latitudes if there
is a thick disc.

\section{Using CMDs to obtain the parameters of the disc}
\label{.1}

\subsection{The selection of giants}
\label{.selecgig}

\begin{figure}[!h]
{\par\centering \resizebox*{6.7cm}{6.7cm}{\includegraphics{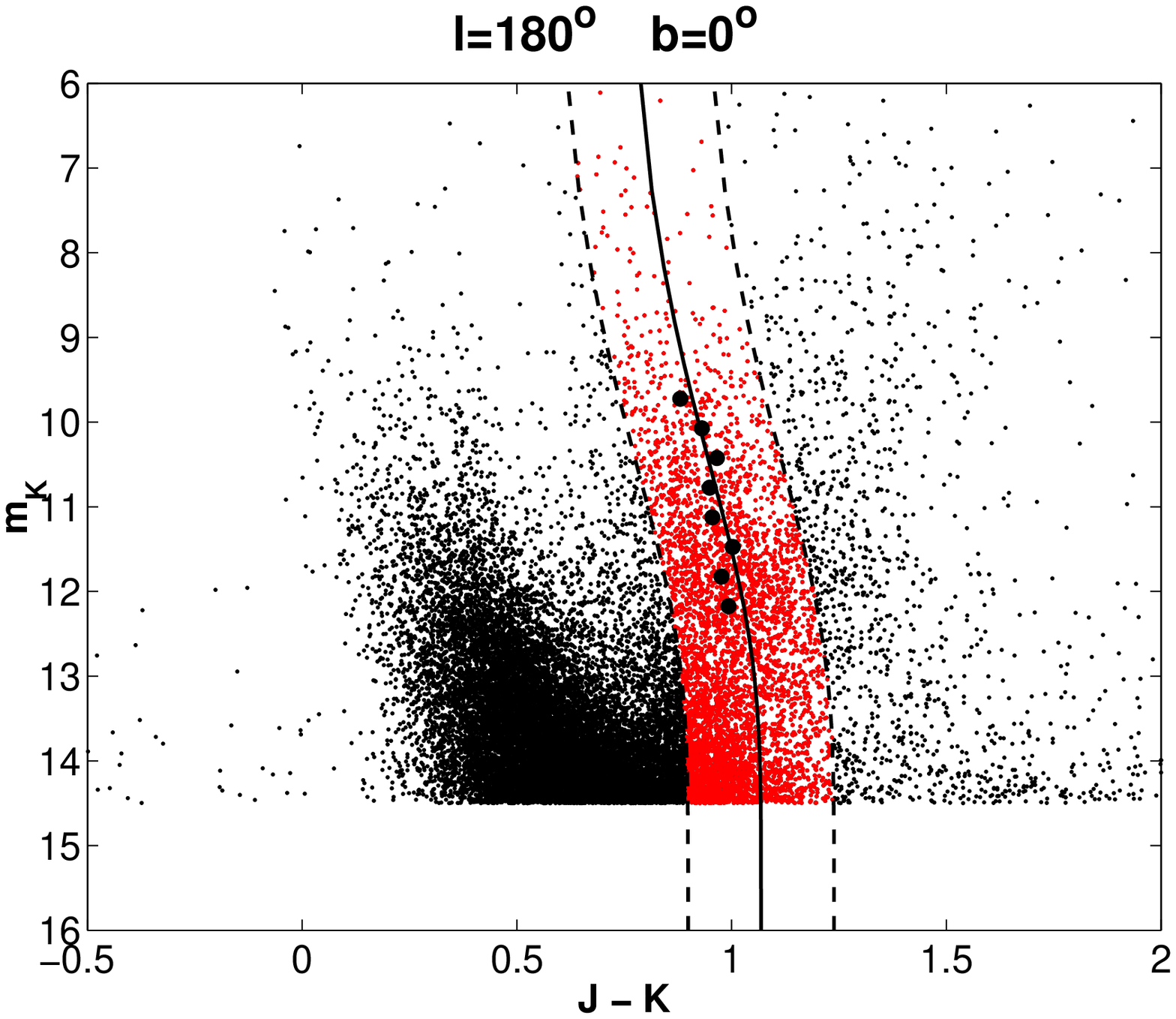}}
\resizebox*{6.7cm}{6.7cm}{\includegraphics{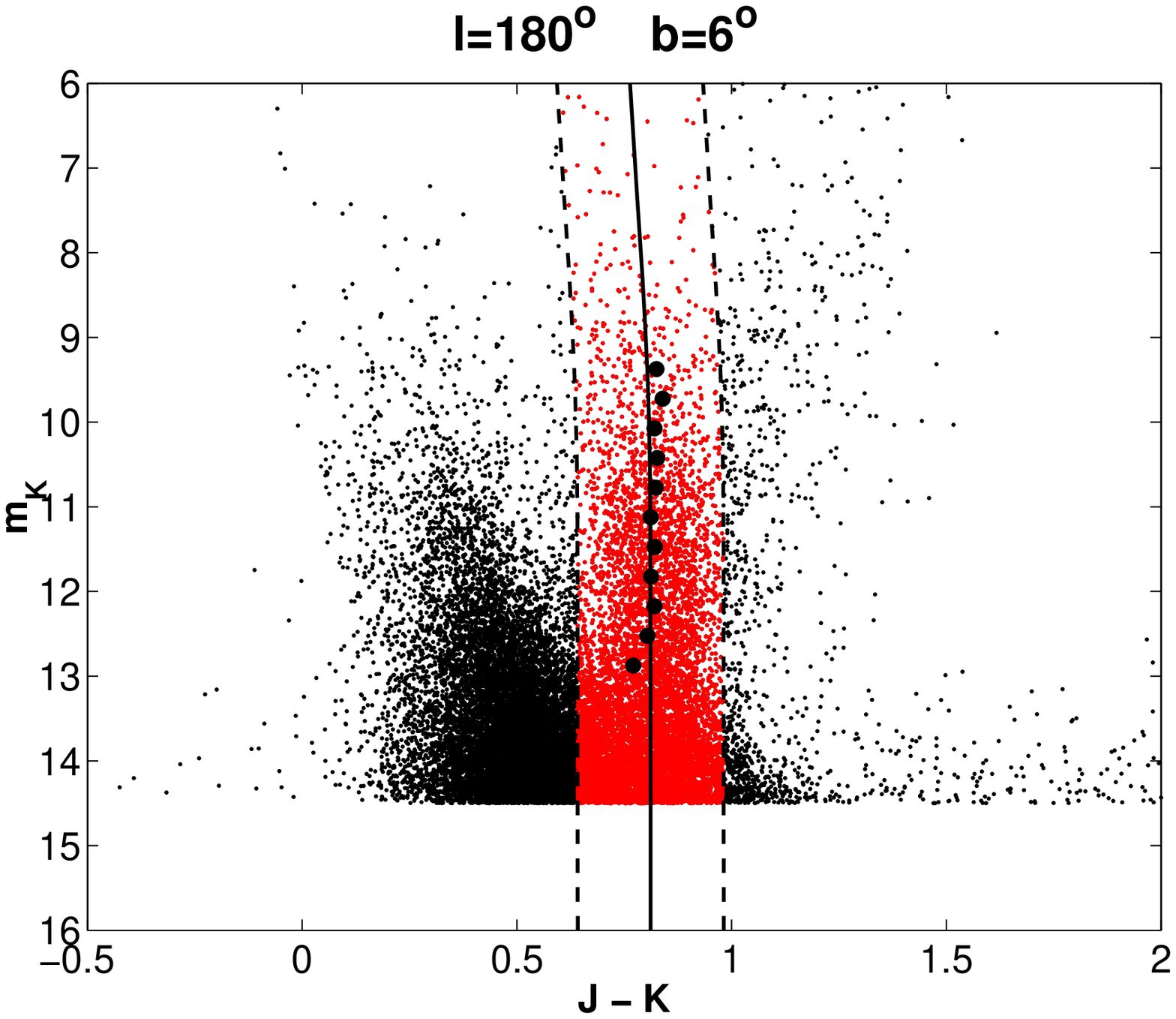}}\par}
\caption{Color-magnitude diagrams for two studied fields. The higher extinction 
causes a slight broadening in the K-giants strip for in-plane regions.
Maxima obtained by a gaussian fitting to star distribution in the
red clump strip. The solid line shows the fitted trace that we assign to red clump giants
population. The dashed lines show the limits for the red clump giants
extraction within a width of 0.4 mag.}
\label{Fig:CM2}
\end{figure}

We have built ($J-K,m_K$) CMDs from  the 2MASS data set for selected regions 
of the sky (Fig. \ref{Fig:CM2}). We have taken the data `as is',  without further
processing. The data have been grouped in regions covering between 2 and 5
squared degrees on the sky (see \S \ref{.selreg} for details). 

The position on a CMD  of a star is determined from its absolute magnitude,  intrinsic color, distance and
extinction. Any horizontal displacement is caused by extinction alone, 
and hence if the intrinsic color of the source is known, the
total extinction  can be directly determined from the measured color.
The vertical motion, however, is caused by both distance (doubling the distance
makes the source 1.5 mags fainter) and extinction. So if the absolute magnitude
of the source is already known, the extinction can be determined from the
reddening and hence the distance is directly determined (see formulae
in \S \ref{.sdens}).
When there are a number stars of the same spectral class (i.e  approximately the same absolute magnitude
and color) but at different distances from the Sun, these will be situated
along a line on the CMD. Increasing the distance shifts the star
to a fainter magnitude, while extinction by itself shifts the stars diagonally.
Hence, combining these two effects would mean that sources from a single 
spectral class will form a line (or stripe) on a CMD.  In areas of the Galaxy 
where extinction is low this line will be almost vertical whereas for lines 
of sight into the inner Galaxy,  the line will have a considerable angle as 
the extinction becomes the dominant effect. 

On a color - absolute magnitude diagram (CAMD), the majority of the
stars lie in the main sequence or giant branch. When there is a wide 
range of distances being sampled, each part of the CAMD
produces its own stripe on a CMD and so the overall effect is more 
of a broad smear rather than a well defined line.  By
far the most common type of giant, however, are the early K giants or red clump
giants. These produce a very well defined grouping, with a density far
higher than the surrounding areas, on a visible CAMD (e.g. Hoeg et al
1997). On an IR CAMD (e.g Cohen et al. 2000) the grouping
is even tighter. For $V-[8.3]$ 
CAMD the total spread of the red clump is
about 0.5 in color and 1 magnitude in absolute magnitude. There
are sufficient of these sources to produce a clearly recognizable
feature in a CMD, and this was used in Hammersley et al. (2000) to examine
lines of sight into the inner Galaxy.  The 
luminosity function derived in  Hammersley et al. (2000) shows
that at the peak of the red clump there are about 10 times the numbers
of sources as there are either 0.8 magnitudes  brighter or fainter, and
the sigma in absolute magnitude is about 0.3 magnitudes.  The $M_K$
of the center of the red clump  is $-1.65$, although there is a possible
error of about 0.1 magnitudes. Using the tables in Wainscoat et al.
(1992) an $M_K=-1.65$ corresponds to a K2III and hence the J-K color
is $(J-K)_0=0.75$.

Figure \ref{Fig:CM2} shows the CMD for two of the fields used in this
study. The red clump stripe can be clearly seen. In order to isolate 
the red clump sources, we have made histograms of horizontal cuts (i.e.
in color) through the CMDs, varying the $m_K$ range used.  A gaussian
was then fitted to histogram (i.e. the region containing the red
clump giants) to determine the position of the peak in each horizontal cut. 
Stars lying well to the left of the maxima are almost
totally dwarfs, as there are very few giants with colors bluer than the
red clump stars. Stars lying to the right will be predominantly the M
giants and AGBs, at least brighter than $m_K$=13.  

In order to  simplify the parameter fitting it will be assumed that
the red clump giants all have  $M_K=-1.65$ (i.e. luminosity function
becomes a Dirac delta function) instead of a more realistic case with
some dispersion (see in \S \ref{.critic} a discussion about this approximation). 
The peak of the histogram is used to trace the red clump giants
empirically. However, it can be seen in Fig. \ref{Fig:CM2} that at
fainter magnitudes the effect of the K dwarf population becomes
significant and swamps the red clump sources. Hence, this limits  the
method to  $m_{K}<13.0$.  Figs. \ref{Fig:CM2} show the position of the
peaks  on each of the CMDs. Using the fitted trace for the red clump
giants extinction along the line of sight for each field can be
determined (see  \S \ref{.sdens}).

The red clump giants can be directly extracted using the fitted trace.
Sources with a $(J-K)$ within 0.2 mag of the center of the fitted
red clump  line were extracted, and the star counts were then obtained. The 
width was determined after a number of trials and is a compromise between 
the necessity to avoid other stars types but include most of the red clump
giants. If too small a width were chosen  then any error in the
determination of the position  peak would lead to a loss of sources at
that point and so reduce the measured density.  Using this width, most
of the red clump giants for low extinction regions will be extracted.
However, where the extinction is somewhat patchy (for
instance in negative latitudes) the strip  broadens and a
significant  number  of stars are lost. Therefore, those regions have been  
excluded from the analysis.

One of the main strengths of this method is that it is empirical and
the only assumption being made is that the absolute magnitudes of all of
the sources being extracted is fixed. If, however, we  had taken a
theoretical trace for the red clump giants using a model and isolated
the stars around this, then any difference between  the predicted and
actual extinction would directly lead to errors in the extracted
counts. Another advantage of this  method is that the spatial 
information is obtained directly from the counts. The position
of the sources can be determined directly from the magnitude and extinction
without the need to resort to models.

The error analysis presented in the following sections will only
include the random errors in the fitting and not the systematic errors
outlined in this section. The error in the determination of the
absolute magnitude is taken to be 0.1 magnitudes, which corresponds to a
distance error of 5\%. The possible error in $(J-K)_0$ only affects
the calculation of the extinction, as the peaks of the
color histograms are derived directly from the data. An error
of 0.1 mag in $(J-K)_0$  would lead to an error of 0.07 mag in the extinction 
[see eq. (\ref{st4})] and hence an error of 3\% in distance.
The errors in assuming that the luminosity is a
delta function are also of the order 5\%. To these should be added the
errors in the 2MASS zero points or the assumed extinction coefficients
although these are likely to be small. The photometric error in 2MASS sources
with $m_K<13$ is $\sim 0.03$ mag., however the error 
is much lower when using the mean magnitude in 
a bin with tens of stars. Similarly, the error in the 2MASS-color is negligible
for of the same reason. Hence, as well as the stated
error there will be a systematic error in the distances, but this
should be under 10\%.

\subsection{Stars density. Disk scale-length}
\label{.sdens}

Using the red clump star counts, we can directly obtain the star
density (stars pc$^{-3}$) by means of the stellar statistics equation:

\begin{equation}
A(m)=w \int_{o}^{\infty} r^{2} D(r)\phi(M)dr
\label{st2}
,\end{equation}
where $A(m)$ is the number of stars per unit area of solid
angle $w$ at $m$ in interval $dm$, and $\phi(M)$ is the luminosity
function. The bins used in this paper have $dm=0.1$ in the range
$m_k=8.5$ to $m_k=13.0$.

As has been already discussed, we have assumed $M_K$=$-$1.65 and an
intrinsic color $(J-K)_{0}$=0.75 for a red clump giants. The extinction
$A_{K}(m_k)$, to a distance
$r$, can be determined  by tracing how  $(J-K)$ of the peak of the red
clump counts changes with  $m_{K}$. The extinction is calculated for
any  $m_{K}$ using the measured (J-K) of the peak, the intrinsic mean
color of the stars in the apparent magnitude bin $(J-K)_{0}$, and the color
excess definition and the interstellar extinction values for $A_{J}$
and $A_{K}$, as given by Rieke \& Lebofsky (1985):

\begin{equation}
A_{K}=\frac{(J-K)-(J-K)_{0}}{1.52}
\label{st4}
\end{equation}

Since we have assumed that red clump giants have an absolute magnitude
of $M_{K}$=$-$1.65,  $\phi(M)$  is replaced by a
delta function. Furthermore, by choosing apparent-magnitude intervals of
0.1 mag ($\delta m$) and the related distance intervals($\delta r$), the star 
density can be directly determined using:

\begin{equation}
r=10^{\frac{m_{K} - M_K + 5 - A_{K}(r)}{5}}
\label{st1}
,\end{equation}
\begin{equation}
D(r)=\frac{A(m)\delta m}{wr^{2}\delta r}
\label{densm1}
\end{equation}

The galactocentric mean distance of the stars per
bin ($R$) is calculated using the 
measured mean distance ($r$) in the following equation:

\begin{equation}
R=\sqrt{R_{0}^{2}+(r\cos b)^{2}-2r\cos bR_{0}\cos l}
\end{equation}

With this method, we are inverting the star counts of the red clump
population to obtain the stellar density along the line of sight. This is a simple and direct method and although some assumptions and approximates are required, these can be justified.
Another method of inversion of star counts is presented in  L\'opez-Corredoira
et al. (2000) 
which  solved for the both density and luminosity function. However, this method  only works for  the bulge  and would not be appropriated for the disc. 

\subsection{Testing the method}
\label{.critic}

In order to to verify that the proposed method is correct a number of 
tests were made:

\subsubsection{The difference between the real red clump distribution 
and a Dirac's delta}

\begin{figure}[!h]
{\par\centering \resizebox*{6.7cm}{6.7cm}{\epsfig{file=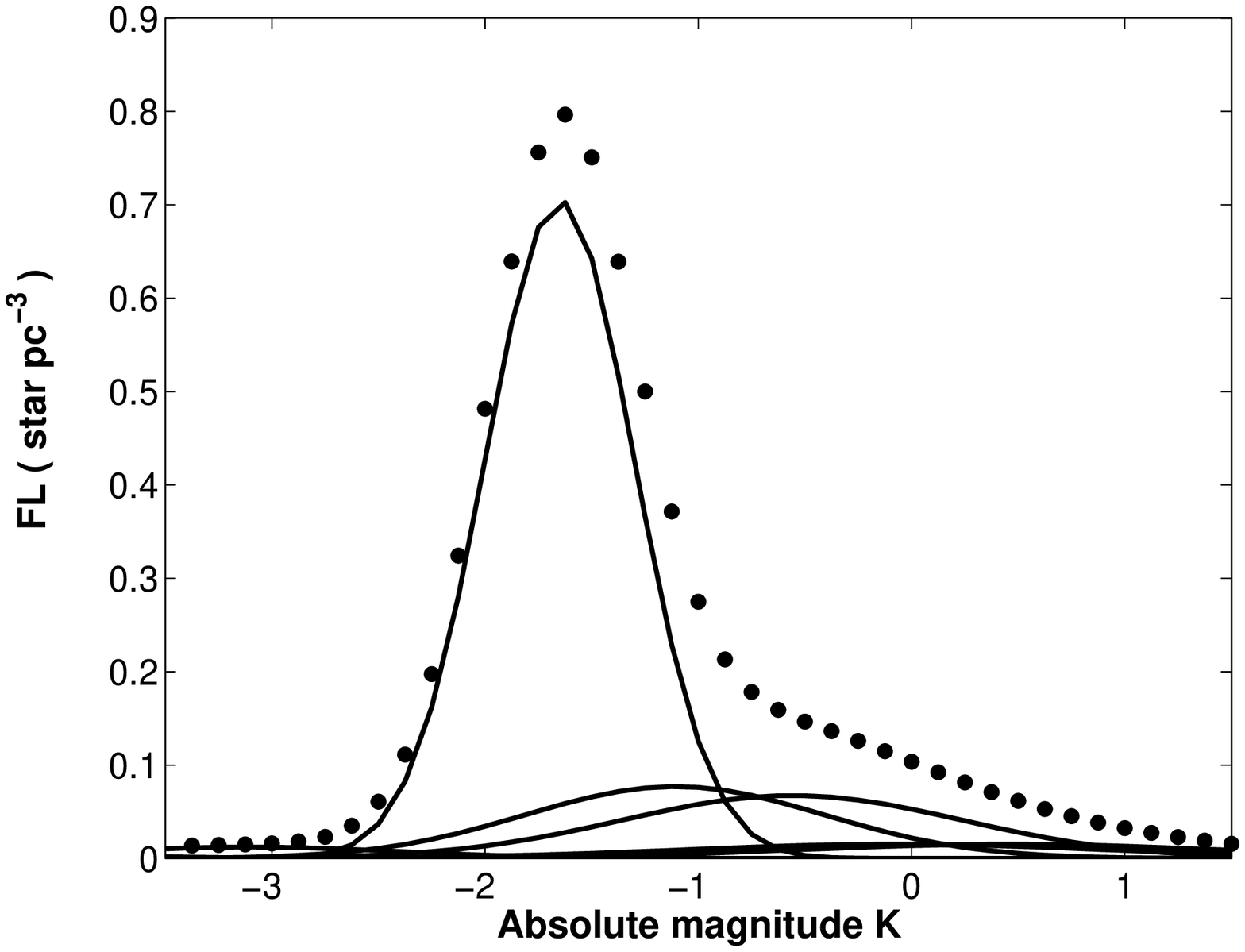,height=6.7cm}}
\resizebox*{6.7cm}{6.7cm}{\epsfig{file=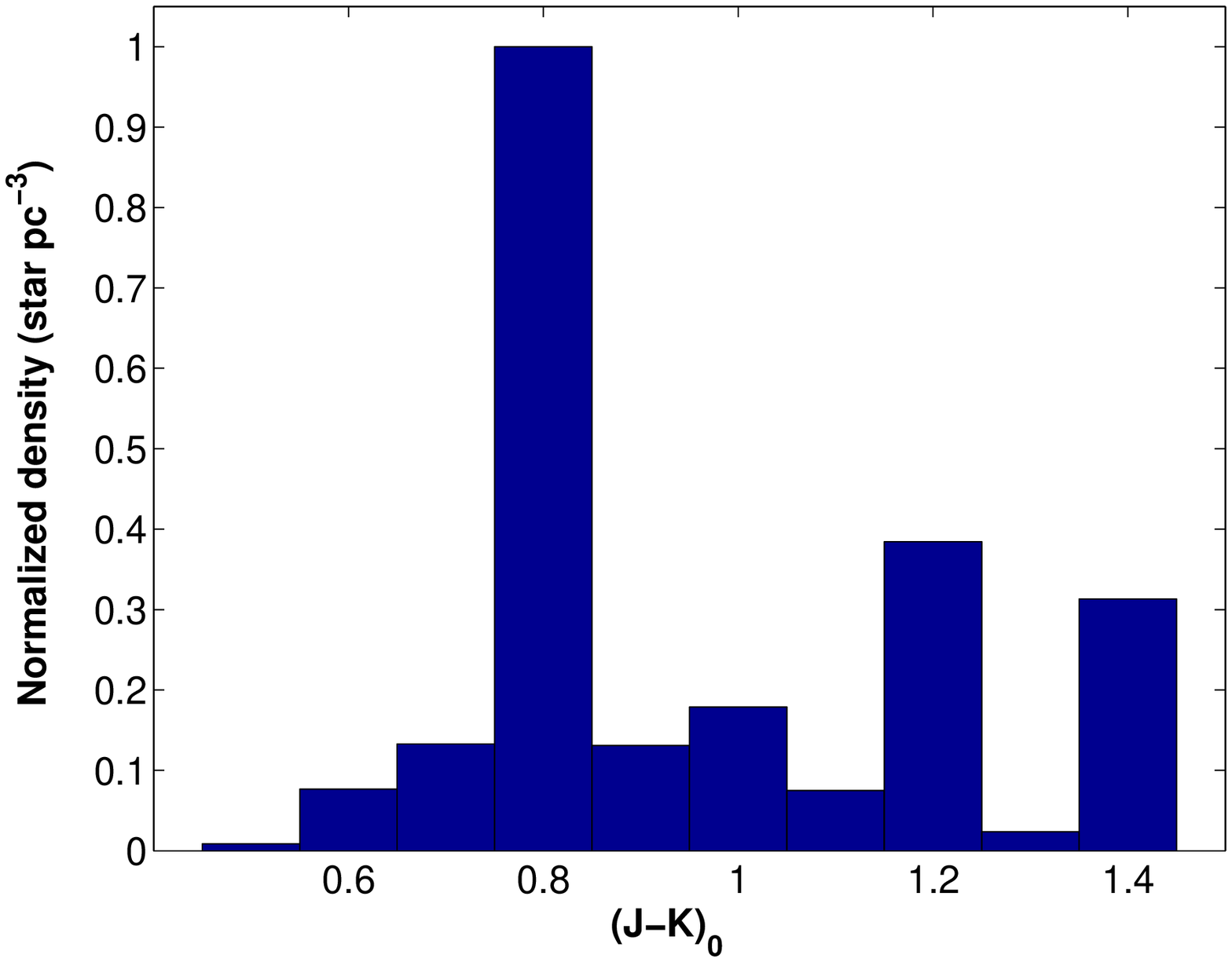,height=6.7cm}}\par}
\caption{Distribution of $K$ absolute magnitudes and $(J-K)_0$ colors
at $m_K=11$ along a line of sight towards the anti-center 
corresponding to the sum of all giant types in the disc according to
the updated ``SKY'' model (Wainscoat et al. 1992; M. Cohen, private
communication). The maxima of these plots corresponds 
the red clump, which are predominant in the giant population.}
\label{Fig:dispK2III}
\end{figure}

To further explore the form of red clump, we have calculated the distribution of
$K$ magnitudes and $(J-K)_0$ colors using the ``SKY'' model,
but updating the density function of the 
giant populations (Wainscoat et al. 1992; Cohen M., private communication). 
In Fig. \ref{Fig:dispK2III}, the part of the luminosity function 
pertaining to the G5 to M0 giants is shown.
The right hand plot shows the predicted form of 
the $(J-K)$ histogram at $m_K$=+11 (but with no extinction) along a 
line of sight towards the anti-center. In both plots
the predominance of the red clump giants is evident. 

\begin{figure}[!h]
\begin{center}
\vspace{1cm}
\mbox{\epsfig{file=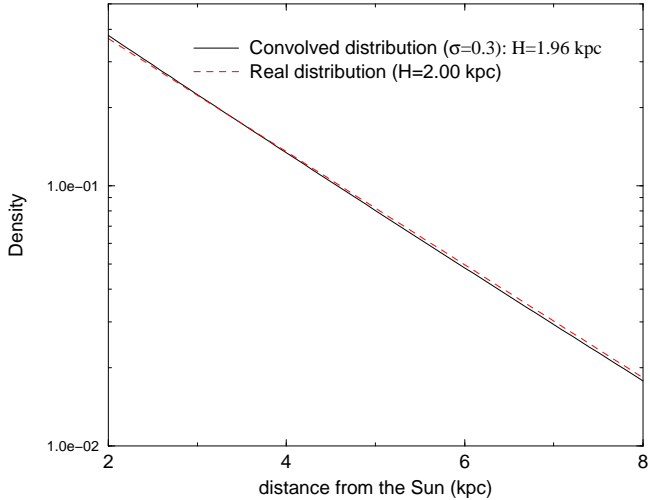,height=6.7cm}}
\end{center}
\caption{Comparison between a real density $\rho=e^{-\frac{d}{H}}$
and the convolved distribution of eqn. (\protect{\ref{conv}}) 
obtained under the assumption of Dirac delta luminosity function
instead of the real gaussian distribution with $\sigma =0.3$.}
\label{Fig:convolucion}
\end{figure}

In order to determine the effect of using a Dirac delta as
the luminosity function for the red clump,
we have carried out a simple calculation. The red
clump giants are assumed to have a gaussian dispersion in magnitude with 
$\sigma=0.3$.
From eqs. (\ref{st2}) and (\ref{densm1}), 
the density in the direction of the anticenter can then be written as:
\[
\rho '(R-R_\odot)=\frac{5}{\ln (10)\sqrt{2\pi }\sigma}
\frac{1}{(R-R_\odot)^3}
\]\begin{equation}\times 
\int _0^\infty dr\ r^2
\rho (r)e^{-\frac{1}{2}\left(\frac{5}{\sigma }\log
\frac{(R-R_\odot)}{r}\right)^2}
\label{conv}
,\end{equation}
where $\rho '$ is the density derived with the method developed in this
section (see \S \ref{.sdens}), while $\rho $ is the real
density distribution. Using eqn. (\ref{roplano}) with
$H=2.0$ kpc, $\rho_\odot=1$, gives the result plotted in Fig.
\ref{Fig:convolucion}. The difference between the two curves is the 
error introduced by the approximation, hence it is negligible.
 The scale-length that is determined assuming  $M_K=-1.65$ for all of the
 sources is $H'=1.96$ pc, i.e. the error
in the calculation of the scale-length is around 2\%. 
Therefore,  the simplification is justified and it does not produce 
significative differences.

\subsubsection{Can the cut-off be masked by  contamination of 
low luminosity giants with the same color than the red clump?}

One critical aspect of the method is the assumption that there are only
red clump sources being extracted (i.e. $M_K=-1.65$). However, the CAMD
presented in Cohen et al (2000, their Fig. 4), shows that there are
lower luminosity sources with the same color as the red clump. (The
range of $(J-K)$ used for extracting the red clump  corresponds to a
$V-$[8.3] of about  1.2). The magnitude limit for the CAMD is about
$m_{\rm [8.3]}=6.5$ and so in the 100 pc plot of Fig. 4 in
Cohen et al. (2000), the source count is
complete for all absolute magnitudes brighter than 1.5.  There are
some  higher luminosity sources with the same color and the red clump,
but these are relatively few and can be ignored. However, there are
significant numbers of lower luminosity giants. These 
giants will have $+2>M_K>-1.4$ and could mask a cut off.

\begin{figure}[!h]
\begin{center}
\vspace{1cm}
\mbox{\epsfig{file=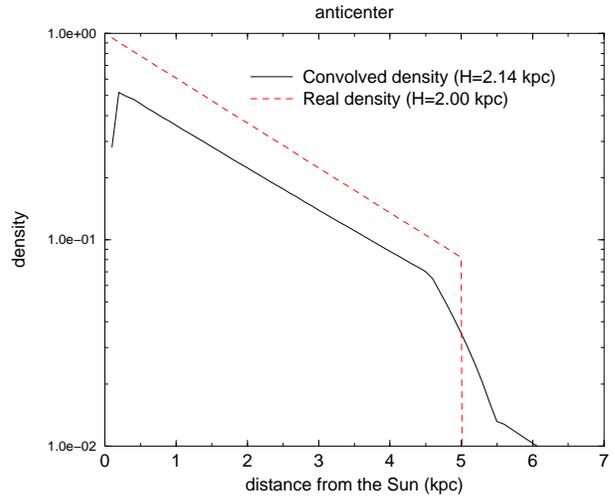,height=6.7cm}}
\end{center}
\caption{Comparison between a real density $\rho=e^{-\frac{d}{H}}$, 
truncated at $R=12.9$ kpc (distance from the Sun: 5 kpc) and
and the convolved distribution
obtained under the assumption of Dirac delta luminosity function
instead of the real distribution with 50\% of the stars with $M_K=-1.65$
and 50\% of the stars between $+2 >M_K>-1.4$ (lower luminosity giants).}
\label{Fig:convolucion2}
\end{figure}

In order to gauge the effect of these sources on the method, a simple luminosity
function was built which had half of the giants with $-1.45>M_K>-1.85$ and
the other half of giants with $+2>M_K>-1.4$.
This is probably an overestimate of the importance of the lower
luminosity sources and so should be considered a worse case scenario.
We have carried out the same type of convolution in the anticenter 
of the previous subsubsection but with this  luminosity function, and we
have used a truncated exponential disc. The result is shown in
\ref{Fig:convolucion2}.

As can be observed, the only significant change is in the amplitude because
half of the stars with $M_K=-1.65$ have been lost. However, this is
not important as the total number of the giants is not used in the
analysis. Also the derived scale-length changes by 7\%. The cut-off,
however, is not masked as there is a clear change of slope in $\log
\rho$ around $d\approx 5 $ kpc. Hence, if there were a cut-off at
about 13~kpc it should  still be visible even with the presence of
possible lower luminosity giants. 
As will it be shown below (Figs. \ref{Fig:gigplano},
\ref{Fig:gig}), we do not observe any change of slope in $\log \rho$, so
the absence of cut-off seems clear and cannot be criticized as an
effect of the contamination of low luminosity giants.

\subsubsection{Dwarf contamination}

\begin{figure}[!h]
{\par\centering \resizebox*{8cm}{8cm}{\includegraphics{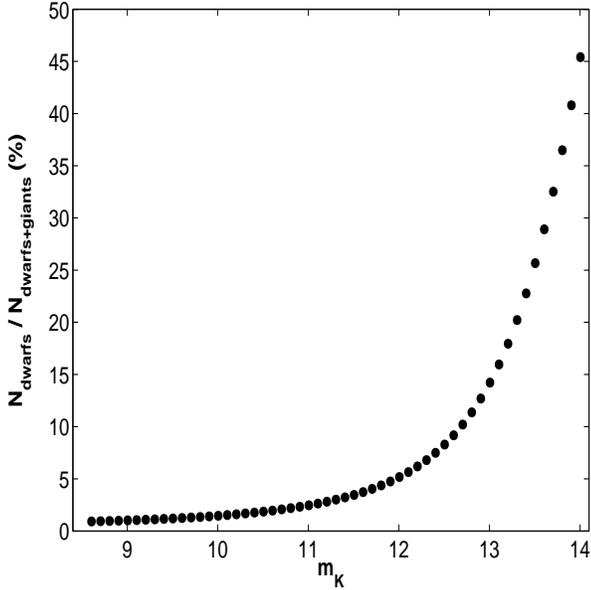}}\par}
\caption{Percentage of predicted dwarf stars in the total star counts extracted from the
color-magnitude diagram for $l=180^{o}$, $b=0^{o}$.}
\label{Fig:dwcont}
\end{figure}

Another critical aspect that must be considered  is the dwarf contamination.
There is an important contribution to the stars counts from dwarf stars for faint 
apparent magnitudes ($m_K>14$; Figs. \ref{Fig:CM2}). The selection method used here  cannot separate
between these two star populations at these magnitudes, as the main
sequence and the K-giant strip overlap. In order to estimate
the dwarf contamination for $m_K<13.0$,  the luminosity and
densities functions of the giants and dwarf stars from  the model of 
Wainscoat et al. (1992) were used to calculate the predicted ratio 
between dwarf and the total of giants plus dwarfs in the counts. The result is shown in Fig.
\ref{Fig:dwcont}. It is clear that for faint apparent magnitudes ($m_{K}\simeq
14$) the dwarf contribution is important,  accounting for almost 50\% of the total counts.
However, the contribution decreases at brighter magnitudes such  that for $m_{K}< 13$ less
than $\sim 10$\% of the stars  are dwarfs. This was calculated for  off-plane regions, where the extinction is almost negligible. It is
expected that for regions close to the Galactic plane, where the extinction is
higher, the dwarf contamination will be even less,
because the reddening will shift the giants to the right whilst the dwarfs will be less affected, and so the dwarf-giant separation is improved. Hence, 
 here we will only use the red clump star counts for $m_{K}<13.0$, where the dwarf contamination is small.

\subsubsection{The ambiguity between reddening and metallicity effects for the
 clump giants}
 
When the metallicity decreases the stars become bluer, and hence the 
reddening could be underestimated.

We have assumed that the stellar populations throughout the disc is homogeneous
and therefore, there are no large gradients in the  metallicity.
However, even in the case of small metallicity variations 
(Cameron 1985, has measured a radial metallicity  gradient of  
-0.11 dex kpc$^{-1}$), the change to the result is small as 
the red clump color is practically independent
of age and metallicity (Sarajedini et al. 1995).
The variations are limited within $\Delta (B-V)_0<0.05$ (Sarajedini 
et al. 1995), which means $\Delta (J-K)_0<0.022$, i.e. an error 
in the extinction less than 0.015, which is negligible.

\subsection{Selected regions}
\label{.selreg}

Twelve fields were chosen from the available 2MASS data,  
which are summarized in Table \ref{Tab:giants}. 
In each field a large area of sky was covered (between 2 and 5 square
degrees) as a large number of stars are needed to determined the
color of the peak of the red clump distribution.  The high number of
red clump giants also minimizes the poissonian noise in the star
counts and so improves the subsequent fitting of the   density
distributions.  The fields were $5^\circ$ in longitude by $1^\circ $ in
latitude, although the actual area covered depends on the available
2MASS data within each region. The star density is changing far
more rapidly in latitude than longitude so it is important not to
include a wide spread of latitudes in each region.  Positive
latitude regions were selected because these are less affected by
patchy extinction, which causes a significant dispersion of the red
clump strips. 
We have only used regions  $150^\circ<|l|<225^\circ$
so as to avoid the northern and southern warps, which  produce 
large asymmetries between positive and negative latitudes (see \S \ref{.warp}).
Furthermore, the lines of sight in $45^\circ <|l|<90^\circ$ travel 
for large distances at approximately
the same galactocentric distance before reaching the outer disc, 
which would increase the error in obtaining the scale-length.

The main characteristics of the chosen fields are summarized in Table 
\ref{Tab:giants}. Column (1) and (2) give
the Galactic coordinates of the center of the field. 
Column (3) is the total area in the field which has 2MASS
data. The total number of stars with $m_{k}<14.5$ is given in column 
(4), and finally column (5) 
gives the total number of red clump giant stars extracted from the 
color-magnitude diagram, with $m_{k}<13$.

\begin{table*}[!htb]
\caption{Selected regions used with the red clump extraction.}
{\centering \begin{tabular}{ccccc}\hline 
\textbf{l (deg)}&
\textbf{b (deg)}&
\textbf{area (deg\( ^{2}) \)}&
\textbf{$N_{stars}$ ($m_{k}<14.5$) }&
\textbf{$N_{{\rm red\ clump}, FWHM=0.4 mag}$ ($m_{K}<13$)}\\
\hline 
180&
0&
3.84&
68492&
2588\\
180&
3&
5&
98473&
4881\\
180&
6&
5&
64366&
3777\\
180&
9&
5&
49614&
7161\\
180&
12&
5&
39017&
5191\\
220&
0&
5&
111788&
5100\\
220&
3&
3.88&
69659&
3756\\
220&
6&
5&
75411&
4024\\
220&
9&
4.99&
54672&
6555\\
220&
12&
3.55&
30867&
3992\\
155&
0&
2.07&
24189&
1225\\
165&
0&
4.22&
71731&
3765\\
\hline
\label{Tab:giants}
\end{tabular}\par}
\end{table*}

\subsection{Results}
\label{.res1}

A sequential fitting procedure has been used to obtain the model parameters
for eqs. (\ref{ro1}) and (\ref{flare}) (or eqs. (\ref{ro2}) and (\ref{H})).
First, to restrict the number of free parameters we have  used only the
data at $b=0$ to obtain the radial scale-length $H$ (i.e., eq. (\ref{roplano})), so that there would
be no dependence on height. This scale-length is then used in the
off-plane regions to derive the values of the scale-height ($h_z(R)$) and flare ($h_{R,flare}$).

\begin{figure}[!h]
\begin{center}
\vspace{1cm}
\mbox{\epsfig{file=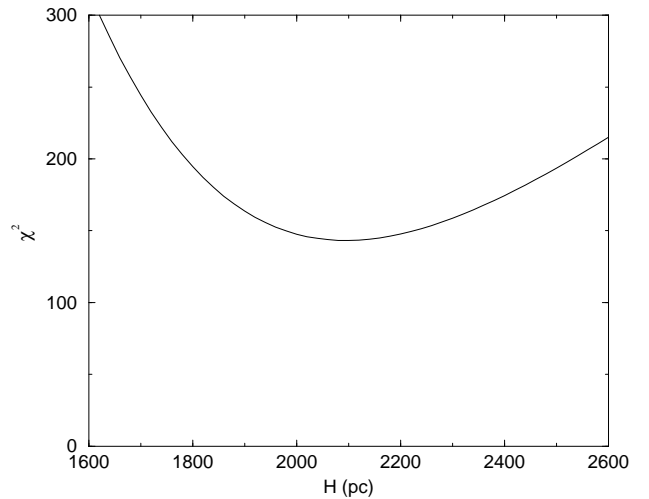,height=6.7cm}}
\end{center}
\caption{$\chi ^2$ vs. $H$ of the fit of red clump star density in the plane of 
the model (\protect{\ref{roplano}}).}
\label{Fig:chigigplano}
\end{figure}

An exponential disc (eqn. \ref{roplano}) was fitted 
to the data with $m_K\le 13.0$ in the plane  
for longitudes of $155^\circ $, $165^\circ $, $180^\circ $ 
and $220^\circ $. The value of $H$ which gives the 
minimum $\chi^2$ ($\chi ^2=143.2$ for a number of data points
$N=184$, see Fig. \ref{Fig:chigigplano}) and 
assuming that the errors for each data point are poissonian
was for:

\begin{equation}
H=2.10^{+0.22}_{-0.17}\ {\rm kpc}
\label{H1}
.\end{equation}

\begin{figure}[!h]
\begin{center}
\vspace{1cm}
\mbox{\epsfig{file=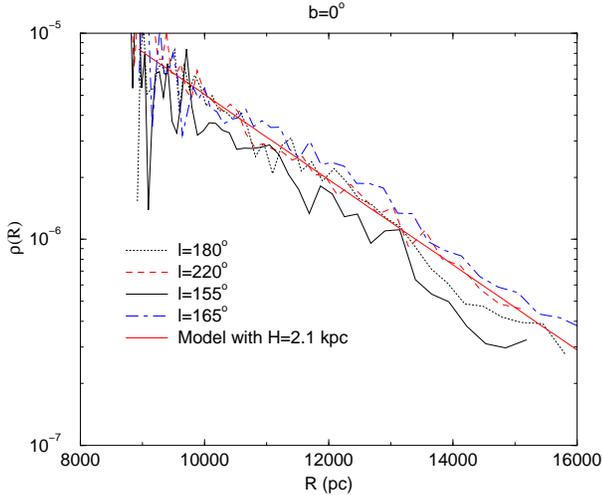,height=6.7cm}}
\end{center}
\caption{Fit of red clump star density in the plane of 
the model (\protect{\ref{roplano}}).}
\label{Fig:gigplano}
\end{figure}

As described in \S \ref{.selecgig}, the red clump sources are extracted from
the general star counts by means of the ($J-K$) CMD. Inserting these number
counts in eq. (\ref{rosol}), with the help of eq. (\ref{roplano}) with the
calculated value for $H$, we derive a density for these sources in the solar
neighborhood $\rho_{\odot, {\rm red\ clump}}=1.31\times 10^{-5}$ stars
pc$^{-3}$. The disc model with the optimum value of $H$ is shown in Figure
\ref{Fig:gigplano} together with the density distribution of the red clump
stars derived as explained in \S \ref{.sdens}. The small differences in the
density  for the different lines of sight stem presumably from the differences
in the patchiness of the extinction, which spreads out  the red clump giants,
or contamination. Even so, it is noticeable how well the analytical function
matches the data.

Using the value of $H$ determined above with off-plane regions at 
$l=180^\circ $ and $l=220^\circ $, allows the disc scale-height 
and the flare scale-length
to be examined. As before, the disc model of eqns. (\ref{ro2}) and (\ref{H})
are matched to the number counts by minimising the $\chi ^2$ of the fit, being
now the free parameters $h_z(R_\odot)$ and $h_{R,flare}$. The best fit 
($\chi ^2=245.4$ for $N=460$) is for:

\[
h_z(R_\odot)=310^{+60}_{-45}\ {\rm pc}
,\]\begin{equation}
h_{R,flare}=3.4\pm 0.4\ {\rm kpc}
\label{hz1}
.\end{equation}

\begin{figure}[!h]
\begin{center}
\vspace{1cm}
\mbox{\epsfig{file=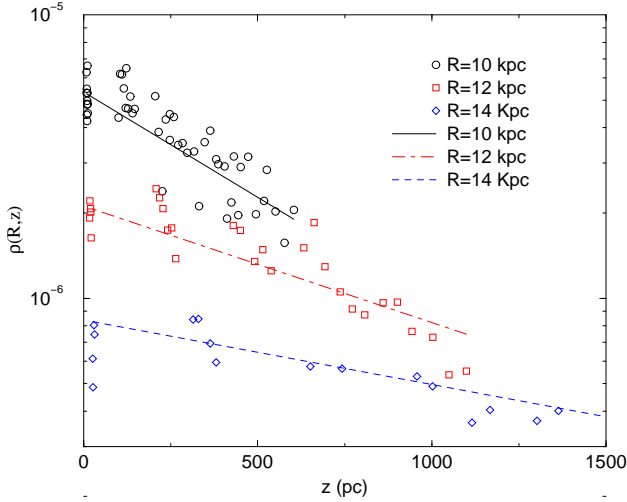,height=6.7cm}}
\end{center}
\caption{The red clump star density obtained from the data and the lines showing a model (\protect{\ref{ro2}}) with $H=2.1$ kpc, $h_z(R_\odot)$=310 pc, 
$h_{R,flare}=3.4$ kpc. Note that the data show a slower decrease of density
for larger $R$, and this fact can be modeled with the flare
$h_z=580$ pc, 1055 pc, 1905 pc respectively for $R=10,12,14$ kpc (the data belong
to a range of 0.5 kpc centered in these galactocentric radii).}
\label{Fig:gigz}
\end{figure}

\begin{figure}[!h]
\begin{center}
\vspace{1cm}
\mbox{\epsfig{file=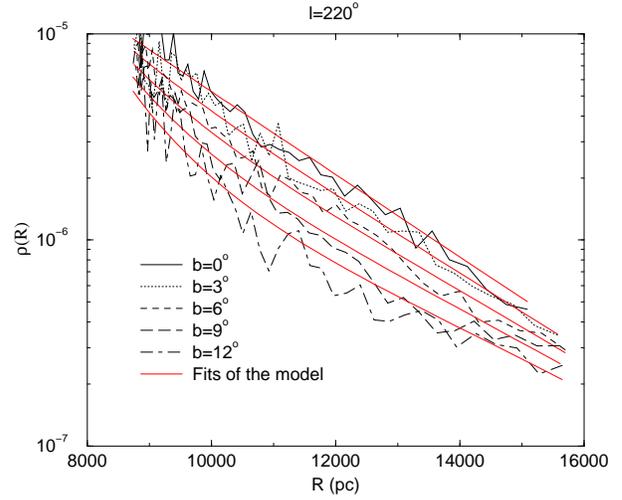,height=6.7cm}}
\end{center}
\caption{The red clump star density obtained from the data and the lines 
showing a model  (\protect{\ref{ro2}}) with $H=2.1$ kpc, $h_z(R_\odot)$=310 pc, 
$h_{R,flare}=3.4$ kpc at $l=220^\circ $ and the latitudes stated on the
plot.}
\label{Fig:gig}
\end{figure}

The errors are calculated for $H=2.10$ kpc and other
values of $H$ could slightly change the error ranges.
The stated errors are determined to be changing the  
parameters such that $\chi ^2$ becomes a factor $f$ larger than the minimum
(which corresponds to a 68\% confidence level and is specified in 
statistics manuals). 
The correlations between the different parameters do not significatively
change the errors.
Figures \ref{Fig:gigz} and \ref{Fig:gig} show the comparison of this 
model with the observational data at different heights above the plane. 
Since the heliocentric distance to the
extracted stars can be measured (see \S \ref{.sdens}), their distance
from the Galactic plane is also known. 
Both figures show that there is an increase in scale-height with increasing  $R$.
Furthermore, in Fig. \ref{Fig:gigz}, the data show a slower decrease of 
stellar density with $z$
at larger $R$ ($h_z=580$ pc, 1055 pc, 1905 pc for $R=10, 12, 14$ kpc respectively).
Both of these effects can only be produced by a flare. 

This result cannot be produced by a combination of a thick disc and a thin
disc with constant scale-heights and the scale-lengths because, then  
the global scale-height would be independent of $R$.
If there is a thick disc with, for example, $h_z=910$ pc, 
a density on the plane of 5.9\% of the thin disc  (Buser et al. 1999) 
and assuming that there is not significative difference in the abundances 
in the red clump population, then  the
thick disc would become dominant for $z>1300$ pc whereas 
for $z<500$ pc the thin disc would be by far the more important.
Fig. \ref{Fig:gigz} shows very  
different slopes for different $R$ at $z<500$ pc,
and even the combination of thin and thick disc
with different scale-lengths cannot reproduce this effect
because the thin disc is highly dominant for $z<500$ pc.
Furthermore, the thick disc cannot
produce significant errors in the calculated scale-length, 
since this was derived in the plane where the thin disc always dominates. 
The scale-height, however, would be representative of an average
of thin and thick discs. The average  source height in the data set used is
$\sim 300$ pc, where the thin disc is $\sim 10$ times denser
than the Buser thick disc.

Another noticeable feature of Figures \ref{Fig:gigplano} and
\ref{Fig:gig} is that the density follows the exponential to at least
15 kpc. As has already been discussed, if the disc had a sharp cut-off then this would cause an abrupt
drop in the density. Hence the absence of such a sharp
drop implies that there is no abrupt cut-off to the disc until at least 
$\sim 15$ kpc. By 15 kpc the star density has become too low to determine 
if there is a cut-off beyond that point. 

\subsection{Flare in the inner Galaxy?}
\label{.inflare}

\begin{figure}[!h]
\begin{center}
\vspace{1cm}
\mbox{\epsfig{file=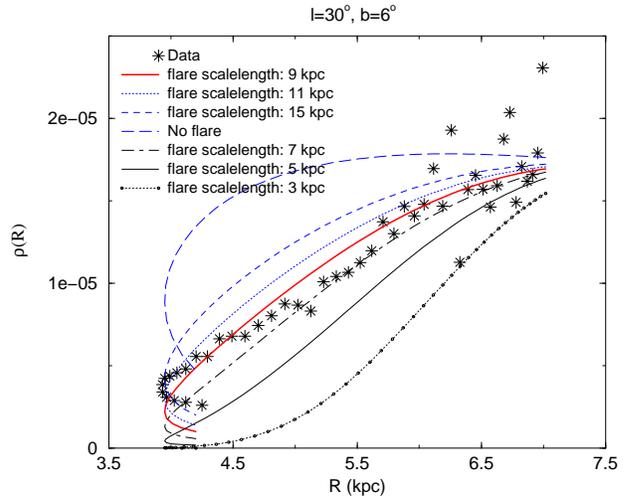,height=6.7cm}}
\end{center}
\caption{The red clump star density obtained from the data and the lines showing a model  (\protect{\ref{ro2}}) with $H=2.1$ kpc, $h_z(R_\odot)$=310 pc
and different flare scale-lengths ($h_{R,flare}$) at 
$l=30^\circ $, $b=6^\circ$. 
As can be seen, a flare is necessary in the inner Galaxy, and
its best scale-length is around 9 kpc.}
\label{Fig:inflare}
\end{figure}

The regions used in the previous subsection all had  
$|l|>90^\circ $, and hence the lines of sight run external to the solar circle
($R>R_\odot$). Therefore, the results are  valid only
for the outer disc. Moreover, the flare 
scale-length of around 3.4 kpc is really only a
description of the flare for $R> 10$ kpc, as between 8 and 10 kpc
there is little data which is also very noisy (Fig. 
\ref{Fig:gigplano}). Hence,
whether this law is applicable to $R<10$ kpc is something that cannot
be deduced from the previous analysis, and data with a  
range of lower galactocentric distances is required.

In order to examine the effect of the flare in the inner Galaxy this
method should not be applied to lines of sight which cover only small a
range  of Galactocentric distances inside the solar circle,
 as these provide only a small range of values with which to attempt
fitting $R$.  For instance, on plane regions with $45^\circ<l<90^\circ$
would only give values of $R$ between 6 and 8 kpc. 
Therefore, in order to explore the inner
disc using this method, we must use regions with $|l|<45^\circ$.
Furthermore, the near plane regions  with $|l|<30^\circ$  should also
be avoided, as the inner Galaxy components become important, e.g.  the
in-plane bar (Hammersley et al. 1994, L\'opez-Corredoira et al. 2001).
At present the available 2MASS data in the regions of interest is very
restricted and so we will concentrate in only one region: $l=30^\circ$,
$b=6^\circ$.

Figure \ref{Fig:inflare} shows the density  of the red clump giants along
this line of sight determined using the same method as in \S \ref{.sdens}.
As well as the measured density, various model predictions are plotted. 
It shows that a flare is necessary to describe
the data with a disc of $H=2.1$ kpc and $h_z(R_\odot)=310$ pc. However, the
scale-length of the flare cannot be as low as 3.4 kpc, but rather around 9 kpc.
The scale-height in the inner Galaxy is smaller than the scale-height
in the solar neighborhood, however the difference is only   $h_z(5\ {\rm kpc})\approx 
0.73h_z(R_\odot)$. A flare scale-length of 3.4 kpc would imply  
$h_z(5\ {\rm kpc})\approx 0.43h_z(R_\odot)$). Hence, although the flare 
 begins well inside the solar circle, the increase of the scale-height with
$R$ is significantly slower in the inner Galaxy.

This result is consistent with previous work. For instance, Kent et al. (1991)
has measured $h_z(5\ {\rm kpc})=0.67h_z(R_\odot)$ by fitting the parameters
of the disc in a 2.4 $\mu $m map of the northern Galactic plane.

\section{Using the star counts to obtain the parameters of the disc}
\label{.scounts}

\begin{figure}[!h]
\begin{center}
\vspace{1cm}
\mbox{\epsfig{file=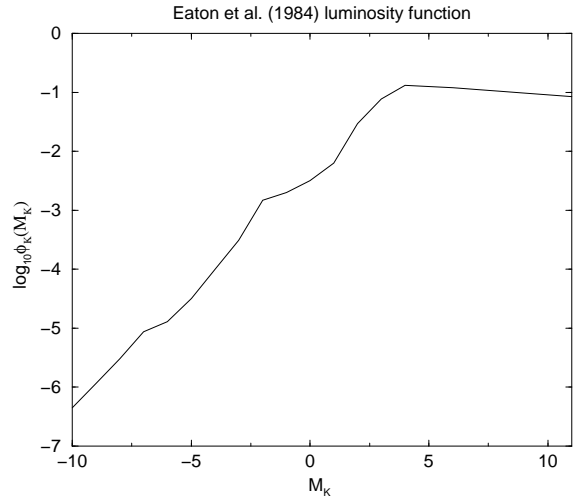,height=6.7cm}}
\end{center}
\caption{K-band normalized 
luminosity function used in the second method of this paper,
from Eaton et al. (1984).}
\label{Fig:PHIeaton}
\end{figure}

The second method used for determining the parameters for the stellar density
is by the fitting of star counts in different regions.
For each of the line of sight centered on Galactic coordinates $(l,b)_i$, 
($i$ is the field number) 
the cumulative star counts ($N_K$) observed in  $K$, 
up to a magnitude $m_K$ in a given  solid angle $\omega $ is
the sum of the stars within the beam brighter than the given apparent magnitude
(Bahcall 1986).
Assuming that the  luminosity function does not vary with the
spatial position, this is 

\[
N_K(m_K,l,b)=
\omega 
\]\begin{equation}\times
\int_0^\infty \Phi_{\rm K} (M_K(r,l,b))
\rho (R(r,l,b),z(r,l,b)) r^2 dr
\label{sc_acum}
,\end{equation}
where
\begin{equation}
M_K(r,l,b)=m_K+5-5\log _{10} r -A_K(r,l,b)
,\end{equation}
\begin{equation}
\Phi _{K}(M_K)=\int _{-\infty}^{M_K}\phi _{K}(M)dM
\label{Phi}
,\end{equation}
$\phi _{K}$ is the normalized luminosity function for the $K$ band in the disc and  
$A_K(r,l,b)$ is the $K$ extinction along the line of sight $(l,b)$ 
up to distance $r$.
We have used a standard 
disc luminosity function given in Eaton et al. (1984, see Fig. 
\ref{Fig:PHIeaton}) and
the extinction  model given by Wainscoat et al. (1992) 
(except with $h_{R,dust}=3.2$ kpc to scale the distance to the center of 
the Galaxy to 7.9 kpc). Fitting could also be used to constrain
the extinction, however it is preferable to have as few free parameters 
as possible. The extinction model of Wainscoat et al. (1992) is axisymmetric and does not include
a flare or a warp in the dust distribution. However,
$K$-band counts are not very sensitive to small variations of the
extinction and were extinction the goal of the paper shorter wavelengths
would be more appropriate. Therefore, as the lines of sight are in general 
looking towards the outer Galaxy and the extinction is low, a 
simple extinction model will suffice.  
 
The stellar density ($\rho $) parameters (eqn. \ref{ro2}) were determined using the
available 2MASS data at  $b=0,\pm 3, \pm 6, \pm 9$ degrees; $45^\circ
<l<315^\circ $. On or near-plane regions were preferred because these
are more sensitive to the variations in the parameters, and there are
significantly more stars.  Again the central regions of the Galaxy were
avoided so that other components other than the disc would not be
included. It should be noted that even when the local stars dominate the counts there are still sufficient giants to constrain the large scale structure.
 Each region used an area of  1 square degree ($\Delta
l=1^\circ $, $\Delta b=1^\circ $). However, some areas are only  partially
covered and in these cases the counts were divided by the actual area
available to obtain the counts per unit area. In total 820 lines of
sight with areas between 0.5 and 1.0 deg$^2$ are used.  This gives a
very large area coverage along the Galactic plane, and so can 
constrain the parameters of the disc. Other authors have used 2MASS
star counts to study the disc however the number of regions used is much
lower. Ojha (2001) used seven independent regions quite far from the plane
but does not study the warp, flare and cut-off whilst 
Alard (2000) only looked at three strips in latitude across the plane.

As noted, this method assumes that all sources have the same spatial
distribution, i.e. the scale-height and scale-lengths in a given
galactocentric radius are the same. For
the scale-height, in particular, this is known not to be true with the
younger components having a scale-height  far less than the older
components. However, the young components only make a significant
contribution to the counts at the brightest magnitudes ($m_K<8$) and
even then only on the plane towards the inner Galaxy (Hammersley et
al.  1999).  For this study we have used
2MASS counts up to $m_K$ = 14.0. This magnitude limit, coupled 
with using lines of sight away from the inner Galaxy, means that the
counts will be completely dominated by the old population of 
the disc rather than the young disc and spiral arms. Hence the
assumption that all of the sources will have, at least, 
very similar scale-heights is justified. 

We have limited the maximum apparent magnitude in $K$ band to 14.0.
This is well below the limiting magnitude of 2MASS and  ensures that
there are  complete statistics in all regions.  Moreover, with
this limiting magnitude, stars brighter than $M_K=0$ can be detected
to over 6 kpc which means the results are sensitive to the shape of the
very outer disc. However, one of the curious features of stars counts
is that the fainter the magnitude the closer on average are the sources
being detected.  This is caused by the steeply rising luminosity
function and means that the dwarfs will dominate the fainter counts.
The result is that the red clump giant method, which isolates a
specific group of stars, will provide a better contrast for the more
external part of the disc.

As discussed previously, the ``old disc'' analyzed here includes
all the old stars of the disc within the  selected regions (average latitude
4-5 degrees). If there is a thick disc like the one proposed
by Buser et al. (1999), the mean contribution to the star counts in
the selected regions would be around 10\%, assuming a
thick disc luminosity function similar to that of  the halo in Wainscoat et
al. (1992). 
Hence, the scale-height determined here would be 
representative of an average of the thin and thick discs but with 90\% of the 
sources coming from thin disc.
 
Star counts cannot easily distinguish between the cut-off and the loss
of stars in the plane due to the flare. However, it was shown  in the previous section that 
it  there is no cut-off to at least 15~kpc, and so we will continue
to assume there is not a cut-off.

\subsection{Results}
\label{.results2}

As in \S \ref{.1}, the fitting of the parameters will be in  
two steps. The  first step is to fit $H$ using the on plane regions ($b=0$) and then fit $h_z(R_\odot)$ and $h_{R,{\rm flare}}$ using all of the regions.

\begin{figure}[!h]
\begin{center}
\vspace{1cm}
\mbox{\epsfig{file=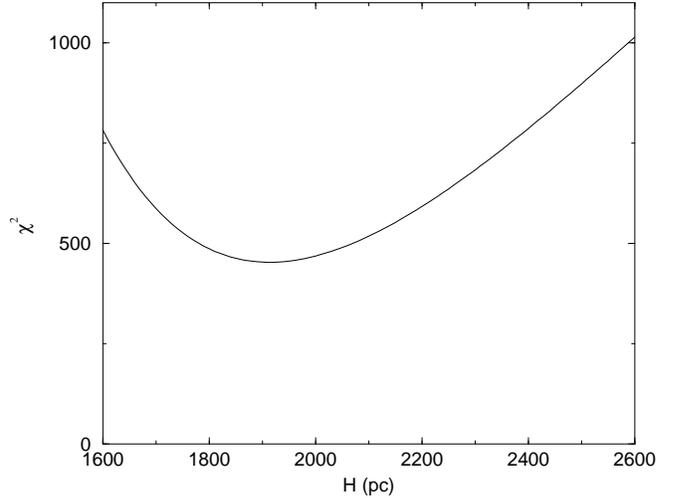,height=6.7cm}}
\end{center}
\caption{$\chi ^2$ of the fit of star counts in the plane for 
the model (\protect{\ref{roplano}}).}
\label{Fig:chiplano}
\end{figure}

Using the available regions in the plane the minimum $\chi^2$ ($\chi ^2=452$, $N=114$) was obtained for  (see Fig.
\ref{Fig:chiplano}):

\begin{equation}
H=1.91^{+0.20}_{-0.16}\ {\rm kpc}
\label{h2}
.\end{equation}
%H=2.15 con Wainscoat/Ortiz; chi2=703
%H=1.73   con Eaton/Ortiz     chi2=518
%H=2.40    con Wainscoat/Wainscoat chi2=541

\begin{figure}[!h]
\begin{center}
\vspace{1cm}
\mbox{\epsfig{file=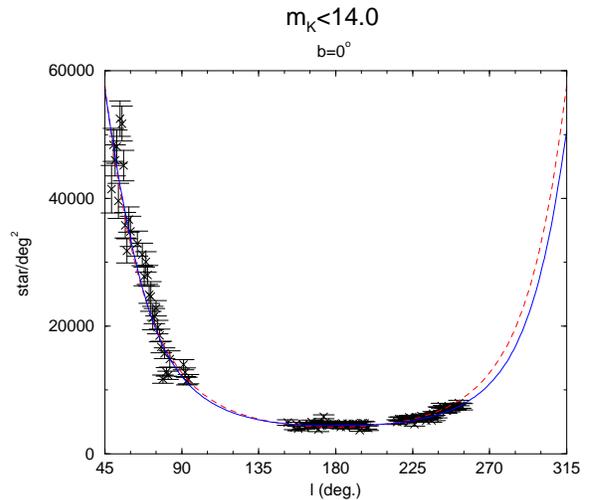,height=6.7cm}}
\end{center}
\caption{Fit of star counts in the plane. The dashed line is
the best non-warped model ($H=1.91$ kpc).
Solid line is the best warped model ($H=1.97$ kpc;
$\epsilon _w=5.25$, $\phi _w=-5^\circ$, $A_w=2.1\times 10^{-19}$ pc).}
\label{Fig:cuen0}
\end{figure}

The stellar density in the solar neighborhood is $\rho_\odot=5.3\times 10^{-2}$
star pc$^{-3}$. This is  somewhat  lower than the value measured in
other surveys (e.g., 0.13 star/pc$^3$, Bahcall 1986). This discrepancy
is most probably due to using different luminosity functions, here we
used the normalized luminosity function by Eaton et al. (1984).
In any case the normalization does not affect the other disc parameters.  The
best fit is shown in the figure \ref{Fig:cuen0} (dashed line).

The other two parameters were then  fitted using  all the available data
at $45^\circ <l<315^\circ $ and $b=0^\circ, \pm 3^\circ, \pm
6^\circ, \pm 9^\circ$. The minimum $\chi ^2$ is found for

\[
h_z(R_\odot)=300^{+13}_{-15}\ {\rm pc}
,\]\begin{equation}
h_{R,flare}=4.6\pm 0.5\ {\rm kpc}
\label{hz2}
.\end{equation}

\begin{figure*}[!h]
\begin{center}
\vspace{1cm}
\mbox{\epsfig{file=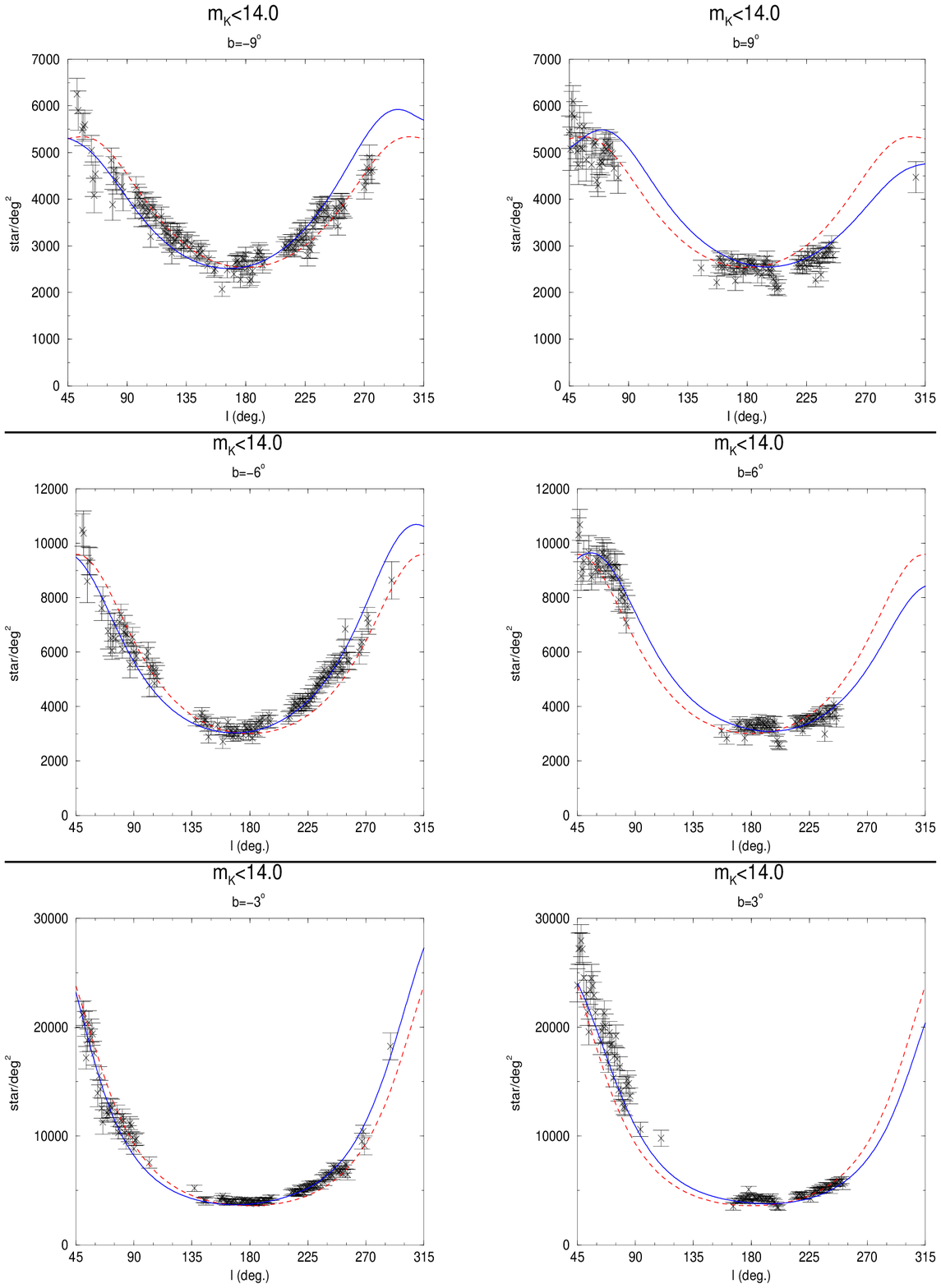,height=18cm}}
\end{center}
\caption{Data and fit of star counts.
The dashed line is
the best non-warped model ($H=1.91$ kpc, $h_z(R_\odot)=300$ pc,
$h_{R,flare}=4.60$ kpc).
The solid line is the best warped model ($H=1.97$ kpc, $h_z(R_\odot)=285$ pc,
$h_{R,flare}=5.00$ kpc;
$\epsilon _w=5.25$, $\phi _w=-5^\circ$, $C_w=2.1\times 10^{-19}$ pc).}
\label{Fig:cuen}
\end{figure*}

These errors are for $H=1.91$ kpc and other
values of $H$ could slightly change  the error ranges.
In this case, the best fit was for  
$\chi ^2=3233$ with $N=820$. The fits are shown in Figures
\ref{Fig:cuen0} and \ref{Fig:cuen} (dashed line).
%The intrinsic scale-length of the disc, without including the effect of
%the flare, would be:
%
%\begin{equation}
%h_R=3.3^{+0.6}_{-0.5}\ {\rm kpc}
%\end{equation}

In general there is  a good fit, although there are some asymmetries
between $45^\circ <l<180^\circ $ and $180^\circ <l<315^\circ $ and also
between $b>0$ and $b<0$. These can be attributed principally  to the
warp (see \S \ref{.warp}), but there are also some
other minor fluctuations, possibly due to the patchiness of the extinction
or contamination from other Galactic components.

\begin{figure}[!h]
\begin{center}
\vspace{1cm}
\mbox{\epsfig{file=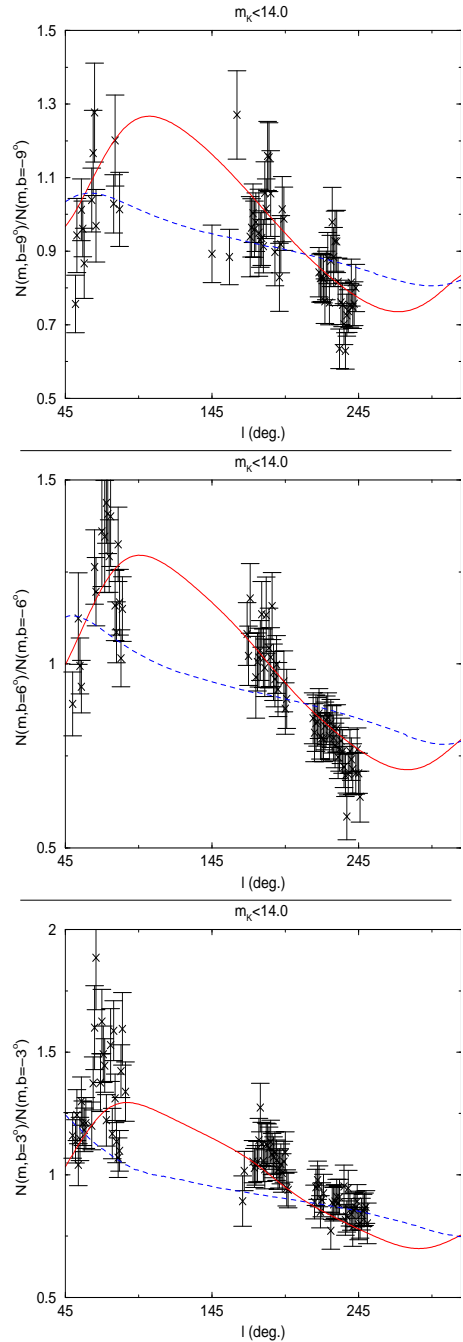,height=18cm}}
\end{center}
\caption{Positive counts divided by the negative counts. The solid line
stands for a warped model with $H=1.91$ kpc, $h_z(R_\odot)=300$ pc,
$h_{R,flare}=4.60$ kpc; $\epsilon _w=5.25$, $\phi _w=-5^\circ$,
$A_w=2.1\times 10^{-19}$ pc. The dashed line stands for the best ``tilted disc'' 
model with $H=1.91$ kpc, $h_z(R_\odot)=300$ pc, $h_{R,flare}=4.60$ kpc;
$b_{\rm corrected}=b+1.64^\circ\sin(l+170^\circ)$.}
\label{Fig:div}
\end{figure}

\section{In search of the stellar warp}
\label{.warp}

If there were neither a warp nor other asymmetries,
the ratio of  counts above and below the plane (Fig. \ref{Fig:div}) should be
nearly the same for all Galactic longitudes, with only a 
small offset due to the height of the Sun above the plane.
However, the plot clearly show a sinusoidal behavior.
There is an excess of counts at positive latitudes with respect the negatives
for $l<180^\circ$ and the opposite for $l>180^\circ $. 
A warp, together with a small offset due the height of the Sun above the
plane, can explain what is seen.

A model similar to  eqn. (\ref{ro2}) will be used, 
but with $|z-z_w|$ instead of
$|z|$, where $z_w$ is the average elevation of the disc above a 
plane which is parallel with $b=0$  but runs though the local center of 
mass (i.e. 15 pc below the Sun).

\begin{equation}
z_w=[C_wR(pc)^{\epsilon _w}\sin (\phi -\phi _w)+15] \ {\rm pc}
\label{warp}
.\end{equation}

Hence, the model  describes the warp as a series of tilted
rings, where the amount of tilt is the galactocentric distance raised 
to a power $\epsilon _w$.  
The  $15$ pc term  is due to the height of the Sun above  the plane
(Hammersley et al. 1995). Since the value of about $15$ pc
is well constrained by other studies we will not carry out an analysis
of this parameter.  The two warp parameters, $\epsilon _w$, and the 
galactocentric angle $\phi _w$ ($\phi _{w \odot}=0$), are determined 
from the division of star counts data in Fig. \ref{Fig:div}.
Once the parameters are fixed then $C_w$ is  obtained from:

\begin{equation}
C_w=\frac{\sum _i\frac{d_it_i}{\sigma _i^2}}
{\sum _i \frac{t_i^2}{\sigma _i^2}}
,\end{equation}
where $d_i=\left(\frac{N(m,+b)}{N(m,-b)}-1\right)$ with N(m,b) the observational
cumulative counts, $\sigma _i$ the errors of $d_i$, and $t_i$ is the model
prediction for $d_i$ with $C_w=1$.

In this way, there are only two free parameters in the model and so it is simpler to constrain the warp. The model assumes that the northern and southern  warp 
are symmetric, apart from the effect of the height of the 
Sun above the plane. This is an approximation and it is known that the gas warp is not symmetric (Burton 1988).  However, the asymmetries of the warp 
are produced over galactocentric distances $R\approx 13$ kpc 
(distance from the Sun: $r>8$ kpc for $l=250^\circ $), 
where there are few stars from the total sample of 
stars with $m_K<14.0$.

\subsection{Results}

\begin{figure}[!h]
\begin{center}
\vspace{1cm}
\mbox{\epsfig{file=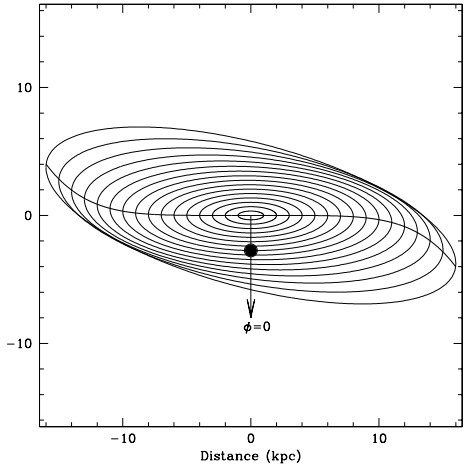,height=6.7cm}}
\end{center}
\caption{A schematic of the warp showing the relative location of the Sun. 
The rings are each kpc, and the line of the maximum warp is shown.}
\label{Fig:warpsch1}
\end{figure}

The best fit is found for:

\[
\epsilon _w=5.25\pm 0.5
%^{+0.6}_{-0.4}
\]\begin{equation}
\phi _w=-5\pm 5 \ {\rm degrees}
,\end{equation}
which gives $C_w=2.1\times 10^{-19}$ pc with $\chi ^2=332.7$, $N=217$.
The best fit is shown in Fig. \ref{Fig:div} (solid line). 
Although there are few points which do not fit very well 
(perhaps due to irregularities in the extinction
or in the distribution, contamination of the spiral arms, etc.) 
again the  general fit is  good. Hence, a warp does accurately
describe the deviations from the best symmetric disc.
In Fig. \ref{Fig:warpsch1}, a representation of the warp is plotted.

\subsection{Could this be a local tilted disc effect?}

An alternative hypothesis that could explain the asymmetries is that
the disc is locally tilted with respect to the global galactic plane.
Hammersley et al. (1995) showed that a similar asymmetry  COBE-DIRBE
surface brightness-maps could be caused by a local tilt to the
disc. However, Hammersley et al. (1995) deliberately chose regions well 
away from the plane, so that the results would be dominated by
local sources and there would be little contribution from regions where
the warp becomes significant ($R>$11~kpc).
Star count data provide direct information on the distance
to the sources, something not available with surface brightness-maps and
so can better distinguish between a tilt and warp.

In Fig. \ref{Fig:div}, it is shown the best fit for a tilted disc
model, i.e. assuming that the asymmetries are only due to
the height of the Sun above the plane and the local tilt of the disc
plane, which corrects the Galactic latitude as (Hammersley et al. 1995):

\[z_w=15\ pc,\]
\begin{equation}
b_{\rm corrected}=b+\alpha_t\sin(l+l_t)
.\end{equation}
The best fit for the tilt corresponds to

\begin{equation}
l_t=170\pm 20\  {\rm deg}
,\end{equation}
which gives a value of $\alpha_t=1.64^\circ$ and $\chi^2=517$ ($N=217$). 
The orientation found is in good agreement with Hammersley et al. (1995), 
but the tilt is some 4 times greater indicating sources at a significantly 
larger distance are being detected. 

The predictions for the tilt are significantly poorer than for the model
of the warp, as can be seen in Fig. \ref{Fig:div} (dashed line).
Hence, whilst a small tilt is sufficient to describe the geometry of local disc,
beyond a few kpc this simplification is no longer valid.
Interestingly the average tilt predicted by the warp model over the
distance range used in Hammersley et al. (1995) (i.e. about 2 kpc) is
still about twice as large  as that measured from the surface
brightness map. This suggests that the warp actually starts just
outside the solar circle, although it should be noted that the predicted
effect in the solar vicinity is small in any case and the star counts
are not sensitive for very short distances from the Sun.

\subsection{Confirmation of the warp with the red clump giants}
\label{.corr1}

\begin{figure}[!h]
\begin{center}
\vspace{1cm}
\mbox{\epsfig{file=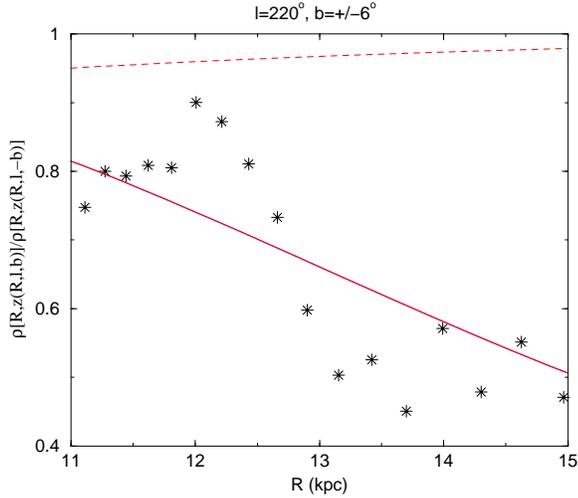,height=6.7cm}}
\end{center}
\caption{Ratio of densities between
positive and negative latitude for ($l=220^\circ $, $b=\pm 6^\circ $).
The solid line shows the predicted ratio for the warp model with $H=1.91$ kpc, 
$h_z(R_\odot)=300$ pc, $h_{R,flare}=4.60$ kpc; $\epsilon _w=5.25$, 
$\phi _w=-5^\circ$, $C_w=2.1\times 10^{-19}$ pc. The dashed line
shows the effect of the height of the Sun above the plane
(+15 pc).}
\label{Fig:divgig}
\end{figure}

A warp should also leave a clear imprint in the distribution of the
red clump giants (\S \ref{.1}).
Since the warp will affect the distant sources most, the asymmetries 
should be greater at larger galactocentric distances.
This can be seen in Fig. \ref{Fig:divgig}, where the ratio of densities at
positive and negative latitude is shown for the region $l=220^\circ, b=\pm 6^\circ$. 
The data show a strong  asymmetry  with most points having a ratio $<0.8$ 
whilst for those beyond 13~kpc it drops to under 0.6.
This again shows that the asymmetries in the counts are larger for 
distant sources.
To confirm that asymmetries cannot be due to the  height of the Sun above 
the plane, the dashed line in  Fig. \ref{Fig:divgig} shows the predicted 
level of this effect. The predicted warp, however, is shown as the solid line 
and this does provide a reasonable fit.

Another aspect which needs to be commented upon is that the measured flare
scale-length may be affected by the warp, even in regions where the warp
is not very conspicuous ($l=220^\circ$). Fig. \ref{Fig:divgig} shows
that the ratio may be as low as 0.6, i.e. an error of around 20\% for each
of the densities at positive and negative latitude.  This is 
one reason why the results from (\ref{hz1}) should be taken with some care.
However, in the second method, the relative contribution  from the very distant
stars ($R>\sim 12$ kpc) is much lower and the effect of the warp is
smaller (compare the results between \S \ref{.results2} and \S \ref{.refit}).
Hence, the parameters derived from the star counts are probably 
more trustworthy than those derived from the red clump giants in \S
\ref{.1}.

\subsection{Coincidence of stellar and gas warp}

\begin{figure}[!h]
\begin{center}
\vspace{1cm}
\mbox{\epsfig{file=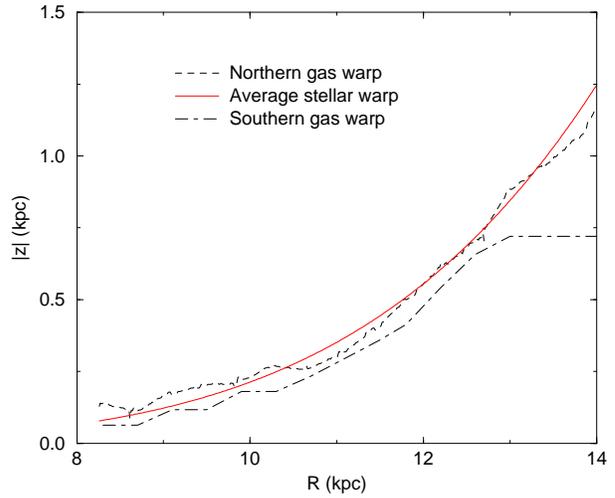,height=6.7cm}}
\end{center}
\caption{Maximum amplitude of the stellar warp (solid line: 
\protect{$z_w=2.1\times 10^{-19} R(pc)^{5.25}$} pc, which is the best power-law
fit to the data) 
in comparison with the one measured by Burton (1988) for the northern and
southern warp gas scaled to $R_\odot=7.9$ kpc (dashed and dot-dashed lines).}
\label{Fig:warp}
\end{figure}

The measurements of stellar warp presented here are totally
independent of those used to determine gas warp (Burton 1988,
Diplas \& Savage 1991). It is noticeable however, 
that the stellar warp is nearly identical to that of the gas. Not
only is the angle of orientation of the maximum amplitude ( 
$\phi \approx 85^\circ$)
coincident,  but the amplitude as well (see Fig. \ref{Fig:warp}). The
northern warp in both the gas and stars are very similar to $R>
13$kpc, as shown in Fig. \ref{Fig:warp}.  The southern gas warp differs
from the northern gas warp, becoming constant or decreasing with height
for $R> 13$kpc. However, since a large majority of the stars along the
line of sight of southern warp ($l\approx 250^\circ $) have $R<13$ kpc
($r<8$ kpc), the data are not sensitive enough to detect the differences
between the northern and southern warp.

Other authors have also looked for the stellar warp. Djorgovski \& Sosin (1989) 
obtained a similar result with IRAS sources.
Alard (2000) also measured the stellar warp using 2MASS data, and
also concluded that the amplitude is roughly coincident with the gas
warp. Although the present results are more accurate, Alard (2000) showed that the amplitude of the stellar warp is consistent with the HI
to within about 30\%.
Drimmel \& Spergel (2001) found from COBE-DIRBE data that the
stellar warp is more or less coincident with the stellar warp found here (and 
Burton(1988) gas warp):
$z_w\approx 25\ {\rm pc/kpc}^2(R-7\ {\rm kpc})^2$ and  Freudenreich (1998) obtained similar results in his independent
determination of warp parameters.

From the presented results,
it is clear that the form of the stellar warp is identical to that of
the gas warp. Hence, whatever is causing of the warp affects both to 
the stars (including the old population) and gas in approximately
the same way. 
Whilst the stellar and gas warps are coincident for the Milky way, 
this may not necessarily be the case in other galaxies.

\subsection{A new fit of the disc including the warp.}
\label{.refit}

After obtaining the parameters of the warp, which is a second order effect on 
the total counts, it is possible update the disc model to include the 
warp and then redetermine the parameters using the cumulative 
star counts. As expected, there are no major
changes with respect to the previous values  (\ref{h2}), (\ref{hz2}):

\[
H=1.97^{+0.15}_{-0.12} \ {\rm kpc}
\]\[
h_{R,flare}=5.0\pm 0.5  \ {\rm kpc}
\]\begin{equation}
h_z(R_\odot)=285^{+8}_{-12} \ {\rm pc}
.\end{equation}
The errors in $h_{R,flare}$ and $h_z(R_\odot)$ are 
for the fixed value of $H=1.97$ kpc. Other
values of $H$ could change very slightly the error ranges.
This best fit gives  $\rho_\odot=5.5\times 10^{-2}$ and corresponds to
$\chi ^2=352$ for the data in the plane ($N=114$), and
$\chi ^2=2049$ for the whole sample ($N=820$).
The fits are plotted in Fig. \ref{Fig:cuen0} and \ref{Fig:cuen} (solid line).
Using eqn (\ref{H}), this provides a  constraint on the intrinsic 
scale-length of the disc:

\begin{equation}
h_R=3.3^{+0.5}_{-0.4} \ {\rm kpc}
\end{equation}

It should be remembered that these parameters have between determined 
between about $R=6$ and $R=15$ kpc (between 0.75$R_\odot$ and 1.9$R_\odot$). 
No data comes from  $R<6$ kpc as the limit in longitude is  $45^\circ $ and 
beyond 15~kpc the data becomes uncertain.

\section{Comparison of both methods for the flare and  discussion}

In \S \ref{.1},  red clump giants gives values
 compatible with those from  the star counts except for the $h_{R,flare}$, 
which is 2.8$\sigma $ lower (3.4 kpc instead of 5.0 kpc). 
It is unlikely that this is a  separate population with different flare,
because the scale-height at the solar circle is coincident. We suspect
the difference stems from a selection effect. Apart from the
fact the measurement of red clump giants is affected by the warp and hence 
the value of 3.4  is somewhat less trustworthy (\S \ref{.corr1}), it must be remembered  that most 
of the giants are at $R>12$ kpc, i.e. 
are dominated by distant stars. However, the 
 flare scale-length is not constant, it is higher for smaller $R$ ( \S
\ref{.inflare} shows that $h_{R,flare}\approx 9$ kpc between $R=4$ kpc and 
$R=7$ kpc). Hence, a systematic difference between the two values should
be expected.

The flare scale-length can be expressed as:

\begin{equation}
h_{R,flare}\approx 12-0.6R \ {\rm kpc}
,\end{equation}
where $R$ is in kpc.
This  gives the values of 9.0 kpc for $R=5$ kpc (close to  the average 
galactocentric distance of the sources along $l=30^\circ $, $b=6^\circ $); 
5.0 kpc for $R=11.7$ kpc (representative of the range between 6 and 15 kpc); 
and 3.4 kpc for $R=14.3$ kpc (representative of the range occupied
by the distant giants). In the solar
neighborhood, the value of $h_{R,flare}$ would be 7.3~kpc.

In comparison, Alard (2000) has described the flare in 2MASS data by
$h_z=h_z(R_\odot )(1+(0.32\ {\rm kpc}^{-1})(R-R_\odot))$. This is similar
to our results in the range $8<R<12$ kpc.
Hence these results are in agreement with   Alard (2000) and   
the  thickening of the disc explains the rapid drop of the density in 
the outer Galaxy without the need to invoke a cut-off for $R<15$ kpc. 
Indeed, as we have seen in 
\S \ref{.1}, the cut-off at $R<15$ kpc is not only unnecessary but 
inconsistent with the data.

The flare, like the warp, is another characteristic of the gas in the outer
Galaxy. Again the parameters determined for the stars are very similar
to that of the gas (Wouterloot et al. 1990), although the gas scale-height
is always far lower.

\subsection{Application of the disc model to brighter apparent 
magnitudes}

\begin{figure}[!h]
\begin{center}
\vspace{1cm}
\mbox{\epsfig{file=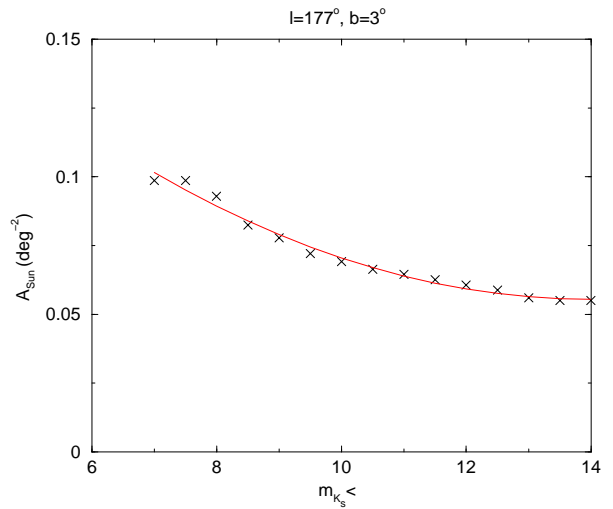,height=6.7cm}}
\end{center}
\caption{Variation of the amplitude $\rho_\odot$ as a function of 
the maximum $K$-band magnitude when we use the disc model of
this paper together with Eaton et al. (1984) luminosity function.
Solid line is eq. (\protect{\ref{ampli}}).}
\label{Fig:amplitud_mag}
\end{figure}

An aspect which must be examined is the possible variation of 
the measured local stellar density  $\rho_\odot$ when using  a range
of magnitudes other than $m_K<14.0$. In principle, the given density
distribution should be independent of the magnitude range used,
however, there may be errors in Eaton's normalized luminosity function that
has been used here. Alternatively there could be source confusion at 
the faintest 2MASS magnitudes, although such an effect would at most  
only affect the sources within a magnitude of the survey limit in the regions measured. 
Both errors would affect the measured 
density, $\rho_\odot$. Figure \ref{Fig:amplitud_mag} shows 
the variation of $\rho_\odot$ with limiting magnitude $m_K<m_{K,max}$. 
This can be described as

\[
\rho_\odot\approx 0.2402-2.642\times 10^{-2}m_{K,max}
\]\begin{equation}
+9.446\times 10^{-4}
m_{K,max}^2
\label{ampli}
,\end{equation}
The data were obtained by normalizing the star counts with a circular
region centered in $(l=177^\circ, b=3^\circ)$ with radius=2 degrees. 
The form of variation is not consistent with there being confusion and so
it has to be due to errors in the form of Eaton's luminosity function.

\begin{figure}[!h]
\begin{center}
\vspace{1cm}
\mbox{\epsfig{file=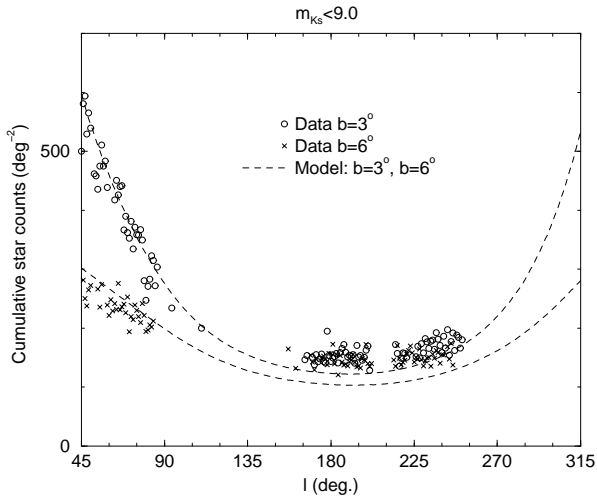,height=6.7cm}}
\end{center}
\caption{Data and fit of star counts for $m_K<9.0$.
The dashed line is the best warped model ($H=1.97$ kpc, $h_z(R_\odot)=285$ pc,
$h_{R,flare}=5.00$ kpc; $\epsilon _w=5.25$, $\phi _w=-5^\circ$, 
$A_w=2.1\times 10^{-19}$ pc).} 
\label{Fig:discpred9}
\end{figure}

An alternative possibility for this variation could have been the density
parameters are not correct and for this reason $\rho_\odot$ is not 
constant with $m_K$. This
can be checked by fitting
of the star counts with the model described in  \S \ref{.refit} but changing the 
limiting magnitude and hence $\rho_\odot$ using eqn. \ref{ampli}. 
It is found that the model still predicts the measured star 
counts (some excess in the anticenter could be due to
some contaminant of bright sources). As an example Fig. 
\ref{Fig:discpred9} shows the fit for the whole plane for $m_K<9.0$. 

\subsection{Application to high latitudes}

\begin{figure}[!h]
{\par\centering \resizebox*{6.7cm}{6.7cm}{\includegraphics{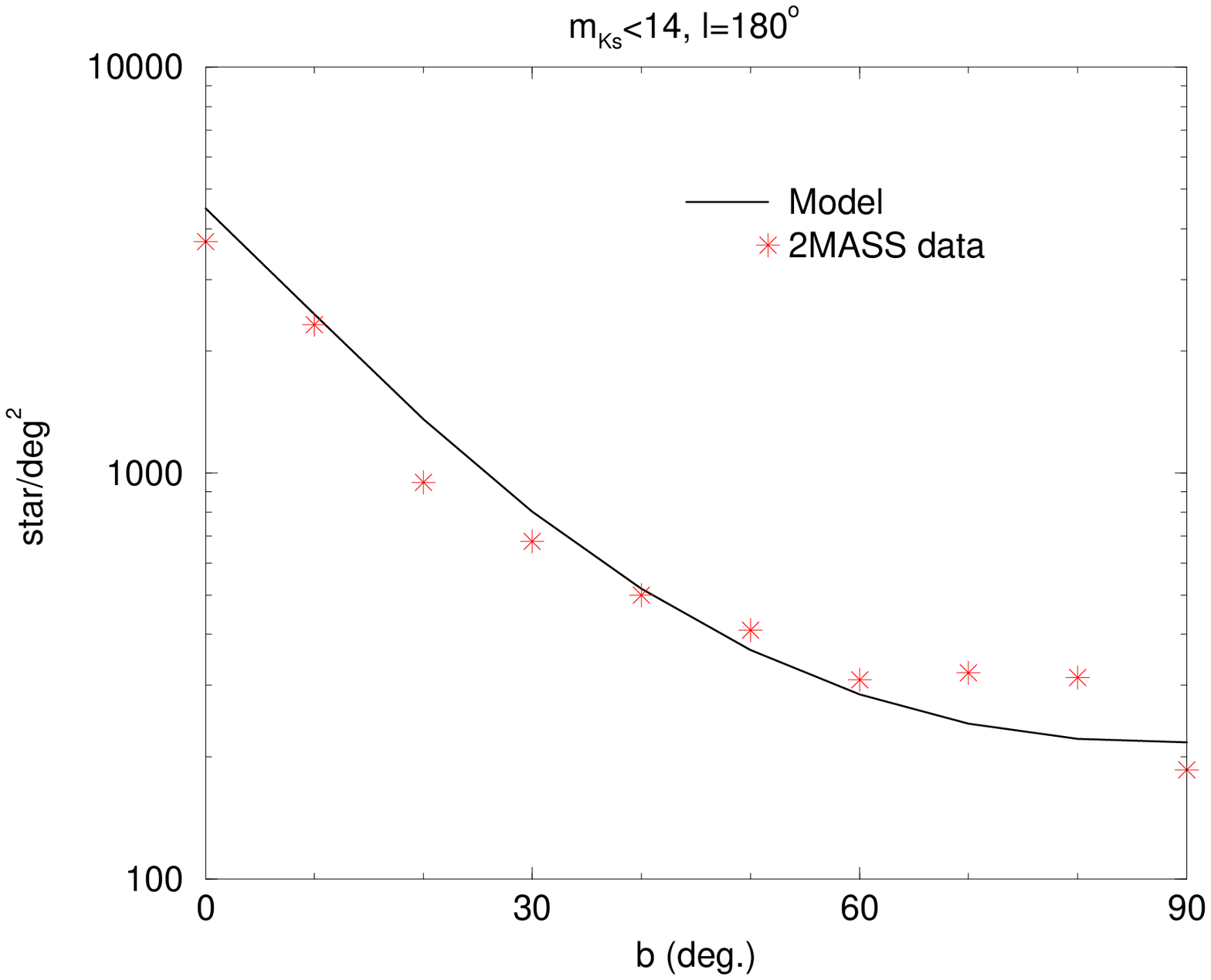}}
\resizebox*{6.7cm}{6.7cm}{\includegraphics{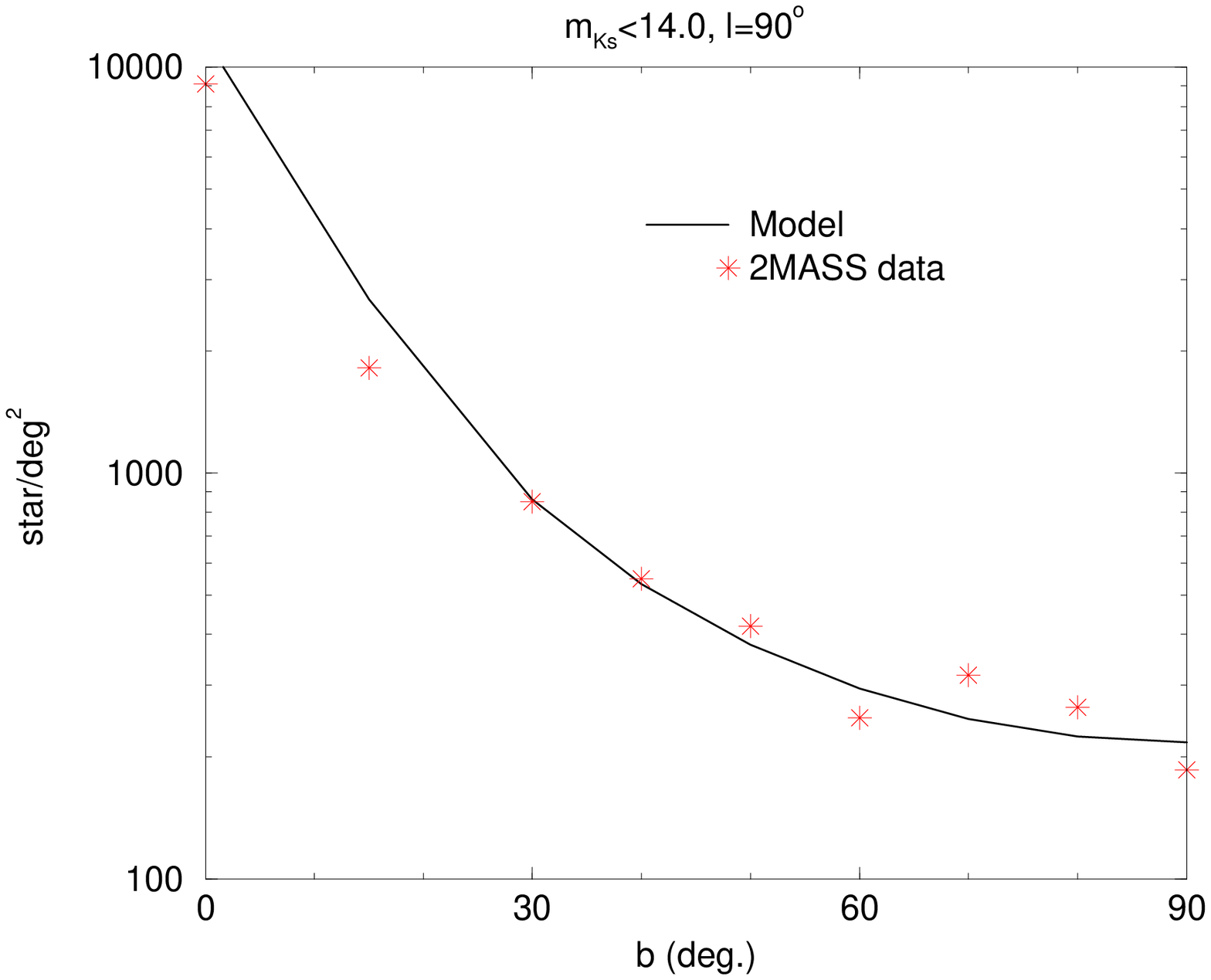}}\par}
\caption{Star counts in high latitude regions. Note how the model derived
in the present paper (fitting low latitude regions) also fit 
approximately these high latitude regions. Data come from the available
covered regions in the different coordinates within a radius of 2 degrees.}
\label{Fig:highlat}
\end{figure}

Fig. \ref{Fig:highlat} shows how well the model fits the counts 
at high latitude regions. 
Even near polar regions ($b=90^\circ $) the model predictions do not 
systematically  deviate  from the counts.  It should be noted that the model does not include a bimodal
distribution, containing  both  a thin and thick disc with  different populations and
scale-heights. None the less  using the thin disc alone the model gives an  accurate prediction of the counts, for the magnitude range being used. 
It is beyond the scope of this paper to discuss whether this result  should be interpreted as an absence of thick disc or whether it is simply that the analysis presented here is not sensitive to the thick disc.  
A more detailed
examination of high galactic latitudes, including an analysis of the
colors and luminosity function, would be required.

\section{Conclusions}

In this paper we examine both quantitatively and qualitatively the
large-scale structures in the Galactic disc. Two independent methods 
are applied to the 2MASS data. The first isolates red clump
giants in a number of regions near the plane, the counts are 
then  inverted to directly obtain the disc density. The second fits the parameters of a disc model  in 820 independent near-plane regions.
Results from both  methods are consistent with each other implying
that both methods are valid.
For instance, magnitude and color of the red clump giants must be close to 
the correct values otherwise there would be significant difference between the 
two methods for  $H$ and $h_z(R_\odot)$.

The distribution of sources within the disc is well described by 
an exponential distribution in both the galactocentric distance and height
above the plane. There is a strong flare, i.e.
an increase of scale-height towards the outer Galaxy. This flare 
starts well inside the solar circle, hence  
there is a decrease of the scale-height towards the inner Galaxy.
This is consistent with the observation that the large majority
of galaxies have flares (de Grijs \& Peletier 1997) and
theoretically consistent with some scenarios of self-graviting stellar
discs (Narajan \& Jog 2002). 
The flare removes the need to invoke a cut-off in the outer Galaxy, 
and certainly there is no evidence for an abrupt cut-off 
in the stellar disc, to at least 
$R<15 $~kpc (=4.5$h_R$). For the analysis it was asssumed that the 
red clump luminosity function could be described as a delta function 
and whilst it is shown that this could have some minor effect on the result, 
it could not mask a cut off.

In other spiral galaxies, the question of the cut-off 
is still not very clear. Pohlen et al. (2002) find that typical radial profiles are better 
described by a two-slope exponential profile, characterized by an inner and 
outer scale-length separated at a break radius, rather than a 
sharply-truncated exponential model. Florido et al. (2001), however, argue 
that  there are some cases of abrupt truncation in the near infrared emission 
of the stellar profile. Kregel et al. (2002) finds that 60\% of
the spiral galaxies are truncated, although he also finds that the
truncation is beyond $4.4h_R$ for galaxies with $h_R<4$ kpc;
so probably, if there existed a cutoff in our Galaxy, it would not
be detected in $R<4.5h_R$.

\begin{figure}[!h]
\begin{center}
\vspace{1cm}
\mbox{\epsfig{file=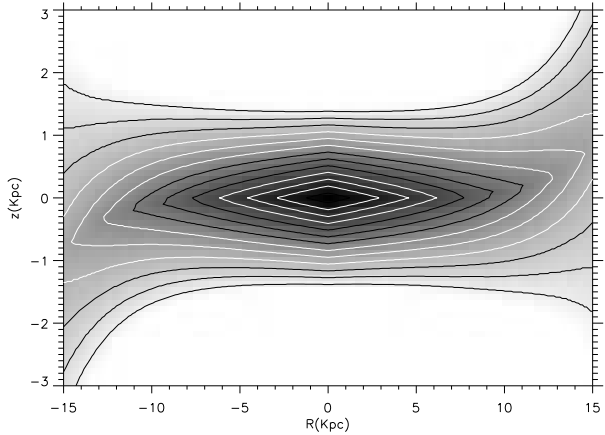,height=6cm}}
\end{center}
\caption{Cut in the plane YZ of the Galaxy as described by the final disc model.}
\label{Fig:figpac1}
\end{figure}

\begin{figure}[!h]
\begin{center}
\vspace{1cm}
\mbox{\epsfig{file=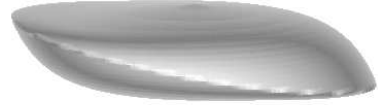,height=7cm}}
\end{center}
\caption{Isodensity surface with $\rho =0.005$ star/pc$^3$ as described by the final disc model. Vertical scale multiplied by
a factor 5.}
\label{Fig:figpac2}
\end{figure}

Another noticeable feature is the warp in the old stellar
population whose amplitude is coincident with the amplitude of the gas warp.
Figs. \ref{Fig:figpac1} and \ref{Fig:figpac2} show the form of 
the disc. S\'anchez-Saavedra et al. (1990, 
2002) or Reshetnikov \& Combes (1998) claim that most spiral galaxies are  
warped at optical wavelengths, but it is not easy to disentangle 
the stellar and gaseous warp.

Various parameters of the old stellar disc have been measured
based on an analysis over  galactocentric distances between 6 kpc and 15 kpc.
The study of the inner disc would be somewhat more complicate,
especially in the plane, because the disc component is no longer
isolated and it is mixed with other possible components: bulge, bar, ring, etc. 
As is shown in L\'opez-Corredoira et al. (2001), the disc might be 
a Freeman type II disc and  inside about 3.5 kpc the disc would be truncated;
this deserves another detailed study which is beyond the scope of this paper.
At present we can talk about the distribution at $R>5-6$ kpc and 
representative of the mean old disc at low latitudes 
(with mean $|z|$ around 300 pc).
If there exists a thick disc, the results for higher latitudes
would be substantially different.
In the following are presented the
results from the star counts, since they are potentially more accurate.
\begin{itemize}

\item The scale-height in the solar circle is $h_z(R_\odot )=285^{+8}_{-12}$ pc.

\item The scale-length of the space density in the plane is
$H=1.97^{+0.15}_{-0.12}$ kpc.

\item the scale-length of the surface density is 
$h_R=3.3^{+0.5}_{-0.4}$ kpc.

\end{itemize}

The errors do not include systematic uncertainties, which we know
to be lower than a 10\%. For the flare and the warp:

\begin{itemize}

\item The variation of the scale-height due to the flare follows 
roughly a law $h_z(R)\approx h_z(R_\odot)
e^{\frac{R-R_\odot}{(12-0.6R(kpc))\ {\rm kpc}}}$ (for $R<\sim 15$ kpc).
\item The warp moves the mid plane of the disc to the height 
$z_w=1.2\times 10^{-3} R(kpc)^{5.25}\sin (\phi +5^\circ )$ pc
(for $R<\sim 13$ kpc), where $\phi $ is the galactocentric 
angle ($\phi _\odot=0$).
\end{itemize}

The results presented here constitute an important step in   
constraining of the disc structure, especially for the stellar 
populations in outer disc, which were previously not very well known. 
Theoretical implications are not discussed here but it is expected that
these results will constitute a valuable test for many
theories about the formation and evolution of the galactic discs, as well
as studies about the formation of warps, flares.

Acknowledgments:
This publication makes use of data products from 2MASS, which is
a joint project of the Univ. of Massachusetts and the Infrared Processing
and Analysis Center, funded by the NASA and the NSF.
We gratefully acknowledge the anonymous referee whose detail report has 
helped improve the paper. Thanks are also given to O. E. Gerhard, 
N. Bissantz and C. Alard for helpful comments.

\end{document}